\documentclass[traditabstract]{aa}
\usepackage{txfonts}
\usepackage{graphicx}
\usepackage{natbib}
\usepackage{longtable}
\usepackage{graphicx}
\usepackage{ulem}
\usepackage{graphics}
\usepackage{keyval}
\usepackage{trig}
\usepackage{graphicx}
\usepackage{dcolumn}
\usepackage{bm}

\def\beq{\begin{equation}}
\def\eeq{\end{equation}}
\def\bey{\begin{eqnarray}}
\def\eey{\end{eqnarray}}
\def\msun{M_\odot}
\def\lsun{L_\odot}
\def\kms{\, {\rm km \, s}^{-1} }

\def\mnras{MNRAS}
\def\apj{ApJ}
\def\nat{Nature}
\def\apjs{ApJ}
\def\apjl{ApJ L.}
\def\araa{ARAA}
	
\def\aap{A \& A}
\def\aj{AJ}

\def\aap{Astron. Astrophys.}
\begin{document}

\title{Implications for dwarf spheroidal mass content from interloper removal}

\author{A.L. Serra, 
          \inst{1,2}\fnmsep\thanks{serra@ph.unito.it}
        G.W. Angus
          \inst{1,2}\fnmsep\thanks{angus@ph.unito.it}
          \and
        A. Diaferio
          \inst{1,2}\fnmsep\thanks{diaferio@ph.unito.it}
          }

   \institute{Dipartimento di Fisica Generale ``Amedeo Avogadro", Universit\`a degli studi di Torino, Via P. Giuria 1, I-10125, Turin, Italy \\
   \and 
             Istituto Nazionale di Fisica Nucleare (INFN), Sezione di Torino, Turin, Italy\\}

\date{\today}

\abstract{Using the caustic method, we identify the member stars of
five dwarf spheroidal (dSph) galaxies
of the Milky Way, the smallest dark matter (DM) dominated systems in the
Universe. After our interloper rejection,
we compute line-of-sight velocity dispersion profiles that are substantially
smoother than previous results. Moreover, two dSphs have
line-of-sight velocity dispersions 20\% smaller than previous
calculations suggested. Our Jeans modelling confirms that the DM content interior to
300~pc is roughly constant with satellite luminosity.
Finally, if we assume that MOND provides the true law of gravity, our
identification of interlopers implies that four dSphs have 
mass-to-light ratios in agreement with stellar population synthesis models, 
whereas Carina
still has a mass-to-light ratio a factor of two too large
and remains a problem for MOND.}

\keywords{ Methods: data analysis --
           Galaxies: dwarf --
           Galaxies: kinematics and dynamics --
           dark matter --
           Gravitation
               }
\maketitle

\section{Introduction}
\protect\label{sec:intr}
Impressive observations by \cite{walker09} and \cite{mateo07} have accurately catalogued
thousands of stars in the vicinity of the eight classical dwarf spheroidal
galaxies (dSphs) of the Milky Way. By binning these line-of-sight (los) velocities
as a function of projected radius and by computing the root mean squared (RMS)
velocity within each bin, it is possible, through Jeans analysis, to reconstruct
the mass profile with considerably less uncertainty than with only global 
los velocity dispersions.

By fitting mass models to these dSphs and to the ultra-faints
(\citealt{belokurov07,koposov07,simon07,willman06}) that were discovered with the aid of the SDSS survey, \cite{strigari08}
have proposed that the mass within 300~pc of every dSph galaxy, over four orders
of magnitude in luminosity, lies in a small range near $10^7\msun$. 
This result is at odds with the naive expectation that more luminous galaxies reside in more massive halos;
one therefore needs to conclude that the \cite{strigari08} result is
of some fundamental significance to the dark matter particle
which creates the halos at high redshift, in which the dSph galaxies form, or
to feedback processes which control star formation (\citealt{koposov09}). 
Using high resolution simulations combined with semianalytical models, \cite{mac09} compute both luminosity and the mass within 300~pc $M_{300}$ of galactic satellites in the $\Lambda$CDM model. They find that $M_{300}\sim10^7\msun$ can be explained by the narrow range of circular velocities covered by visible satellites at the time of accretion.

Several other analyses have been made that rely on the los
velocity dispersions, for instance the calculation of the dSph mass-to-light
ratios $M/L$ (the only truly free parameter) in Modified Newtonian Dynamics (MOND) by
\cite{angus08} and the mass modelling performed by \cite{walker09}.

An issue may exist with these studies because they depend crucially on the
dSph (and all the stars within the projected tidal radius) being in virial
equilibrium, except for some field stars which can and have been rejected 
from metallicity cuts by \cite{walker09}. In
general, there is no concrete way to know if a star is bound to the dSph or
not. This is especially true of the cold dark matter (CDM) paradigm, 
since in principle we can have any density profile of dark matter halo which
can simply be increased to accommodate a fast star. Nevertheless, a re-analysis of the SDSS discovered dSphs Segue I 
(\citealt{nied09}) and Hercules (\citealt{aden09}), that have been performed with 
alternative techniques to the one presented here, has demonstrated that neither of these 
two ultra-faint dSphs are as massive as previously claimed, and unbound stars are surprisingly common.

The point is that if the bound stars have a certain RMS velocity and
just a few unbound stars have RMS velocities significantly higher, the
unbound stars can dominate the combined RMS velocity. Since in Newtonian dynamics galaxy mass is proportional to the second power of the velocity dispersion (and to the fourth power in modified Newtonian dynamics) it is absolutely
critical to make sure the stars used in the Jeans analysis are bound.

This is contrary to the mass modelling of spiral galaxies and clusters of
galaxies. For instance, the masses of spiral galaxies are usually measured from
the velocities of the neutral hydrogen gas which is forced to be on circular
orbits and tidal features would leave very obvious signatures. For one thing,
escaped gas along the los would have an entirely different density to the gas
inside the galaxy. Secondly, warps are relatively easy to identify in gas, but
not from the discrete distribution of stars (\citealt{mccon09}).

The same is true for clusters of galaxies when the mass is inferred from the
measured temperature and density of the X-ray emitting plasma. Alternatively,
cluster masses can be measured with gravitational lensing (e.g., \citealt{clowe06})
or the caustic method \citep{diafgel97,diaf99,serra10}, both of which 
are independent of the state of equilibrium. 

Another important consideration is that the central regions of clusters are not tidally
influenced by their surroundings, both because of their mass dominance and the
time scales involved. Similarly, the majority of spirals evolve in the field,
where galaxy-galaxy interaction are relatively rare.
At the opposite
end of the mass spectrum, dSphs are in close proximity to their hosts ($\sim$~100kpc)
and the tidal field is rapidly changing as their orbits are shorter
($\sim$~100Myr). Furthermore, the dSphs have extremely low surface brightnesses
and are very extended.

Since the wrong identification of bound stars can distort the real velocity dispersion of a system, an accurate detection of interlopers is essential. 
A traditional method for removing interlopers consists of eliminating those candidates whose los velocity is greater than 3$\sigma$ from the mean velocity. The process is iterated until the number of remaining stars is stable \citep{yah77}. A less conservative method is the jackknife that removes, in each step, the stars whose elimination causes the largest change in the virial mass estimator. But these methods are restricted to the information given by the los velocity, and they do not explore the combined information coming from the velocity and the position of the stars.

There are several methods available for performing a membership selection in velocity space. \cite{har96} developed another iterative method taking into account that the maximum radial velocity of a member can be estimated if the mass distribution of the system is known. The mass profile is calculated by the application of the virial theorem and the stars whose velocities are greater than the maximum velocity allowed at its radius are excluded from the calculation of the mass profile in the next iteration step. This technique has been successfully tested  by  \cite{kli07} in simulated dwarf spheroidal galaxies, and its application to real dwarf galaxies (the same ones as studied here) by \cite{lok09} resulted in the removal of significant numbers of potential interlopers and enabled them to find consistent mass-follows-light models that fit the data.

An alternative to these methods is the caustic technique which
utilises the three known phase-space coordinates to delineate caustics (escape speed curves) in the
los velocity-projected radius diagram which enclose the bound stars.
Although the caustic technique was originally designed to study the mass profiles of
galaxy clusters beyond the virial radius (for a review see \cite{diaf09}), it can also function as an interloper
identifier. 

This is of particular significance for the ultra-faint dSphs because lack of
stars, greater susceptibility to tides and more severe impact of interlopers 
aggravate the problem. However, in contrast to the ultra-faint dSphs, the classical dSphs 
have a highly statistically significant number of stars, 
which the caustic technique requires to operate effectively. So we focus on them and await more data on the ultra-faint dSphs.

After we have the confirmed stars as members, we recompute the los velocity dispersion
as a function of projected radius and perform our Jeans analysis in both MOND
(to check whether the $M/L$s are consistent with the stellar populations) and in
Newtonian dynamics to investigate the claim that the mass within a radius of
300 pc is similar for every dSph.

We make use of the recently published data collected by \cite{walker09} for
Sculptor, Fornax, Carina and Sextans, as well as Leo I from \cite{mateo07}.\\

\section{The Caustic Method}
\protect\label{sec:caus}
The caustic technique \citep{diaf99} was originally used to identify galaxy members and compute mass profiles in clusters of galaxies, from the central region to beyond the virial radius. This technique has proven to be very effective in estimating the mass profiles of galaxy clusters from N-body simulations, independently of the dynamical state of the cluster.
Considering that the technique can identify members in potentially any system, we have 
modified it to be used for interloper removal in dSphs. The principle is the same, i.e., 
from the velocity diagram (los velocity vs. projected radius from the centre), 
the technique locates the caustic curves which represent the escape velocity of the system. Therefore, the stars within these caustics are members of the system and they can be used to calculate the velocity dispersion profile.

\begin{figure*}
\center
\includegraphics[angle=0,scale=.70, bb=60 100 680 530]{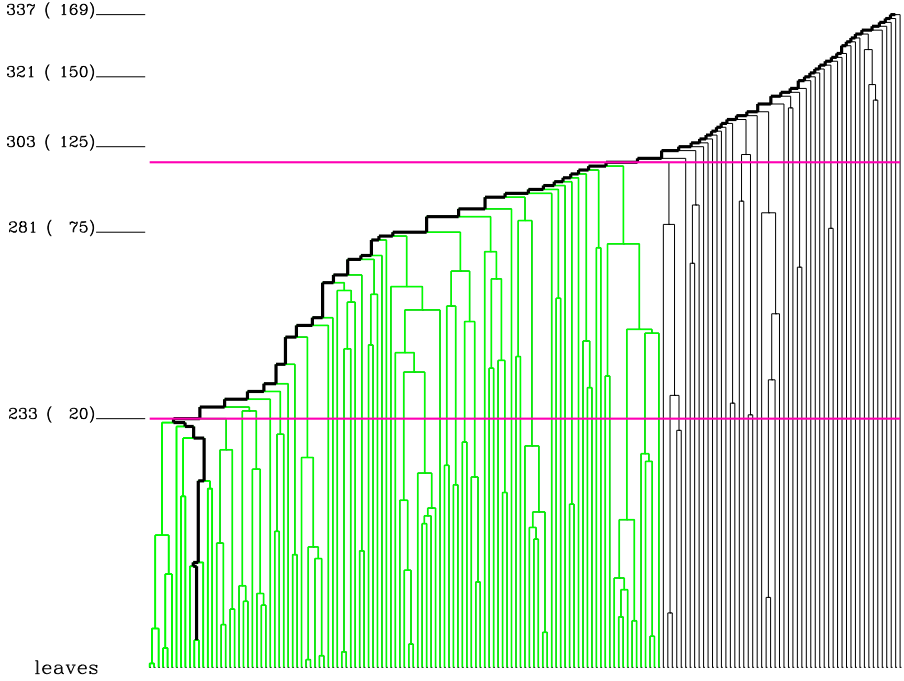}
\caption{\small{Binary tree of a random sample of 200 stars in the field of Leo I. 
Stars are the leaves of the tree (at the bottom). The thick path
highlights the main branch. The horizontal lines show the upper
and lower thresholds used to cut the tree. Some nodes are labelled on the left side, with the 
number of stars hanging from those nodes between parentheses.}} 
\label{fig:dendrogram}
\end{figure*}

In the velocity diagram of an astrophysical system, the components (stars or galaxies) populate a region whose amplitude 
decreases with increasing projected distance $r$ from the centre. 
This amplitude ${\cal A}$ was identified by \cite{diafgel97} to be equivalent to the escape velocity along the line of sight, i.e. ${\cal A}^2(r)=\langle v^2_{\rm esc, los}\rangle$. To locate the caustics, it is necessary to begin by determining a set of candidate member stars of the dSph. These stars will provide a centre, a size and a value for the los velocity dispersion, once they have been arranged in a binary tree (Figure \ref{fig:dendrogram}), according to their pairwise ``projected'' binding 
energy 
\begin{center}
\begin{equation}
E_{ij}=-G{m_i m_j\over R_p}+{1\over 2}{m_i m_j\over m_i+m_j}(\Delta v)^2 \; .
\label{eq:pairwise-energy}
\end{equation}
\end{center}
In this expression, $R_p$ is the projected separation on the sky between the stars $i$ and $j$, $m_{j,i}$ are their masses, $\Delta v$ is their los velocity difference and $G$ is the gravitational constant.

The binary tree is built as follows: (i) each star can be thought as a group $g_\alpha$, then (ii) at each
group pair $g_\alpha, g_\beta$ is associated the binding energy $E_{\alpha\beta}={\rm min}\{E_{ij}\}$, 
where $E_{ij}$ is the binding energy between the star $i\in g_\alpha$ and the star $j\in g_\beta$;
(iii) the two groups with the smallest binding energy $E_{\alpha\beta}$ are replaced with
a single group $g_\gamma$ and the total number of groups is decreased by one; (iv) the
procedure is repeated from step (ii) until only one group is left.

Figure \ref{fig:dendrogram} shows the binary tree of a random sample of
stars in Leo I. The branches of the tree 
connect the root (at the top of the figure) to the leaves, namely the stars
(at the bottom of the figure). The vertical positions of the nodes where the tree branches depart, between the root
and the leaves, depend on the binding energy (\ref{eq:pairwise-energy}) 
of the stars that hang from their branches: the more bound the stars are, the lower the node's location is.

The tree automatically arranges the stars in potentially distinct groups 
with only a single parameter, the star mass,
which enters equation (\ref{eq:pairwise-energy}). 
In the following, we assign the same mass to every star. 
To get effectively distinct groups and to specifically define the set of candidates, we need to cut 
the tree at some level. This level sets the node from which the candidate members hang. 
The level is defined by the ``$\sigma$ plateau'', as described below.

\subsection{The $\sigma$ plateau}

\begin{figure}
\center
\includegraphics[angle=0,scale=0.5, bb=30 20 470 280]{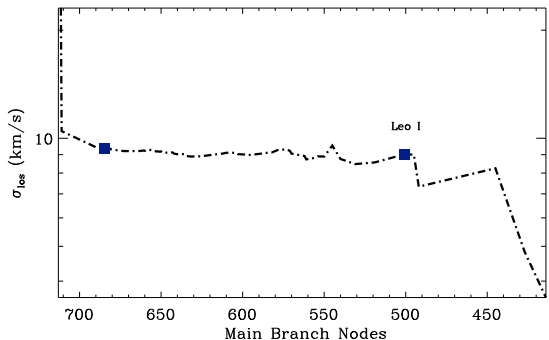}
\vspace{-0.2cm}
\caption{\small{Line-of-sight velocity dispersion of the ``leaves'' (stars) of each node
along the main branch of the binary tree shown in Figure \ref{fig:dendrogram}. 
Filled squares indicate the nodes $x_1$ (left) and $x_2$ (right).}} 
\label{fig:thresholds}
\end{figure}

We identify the main branch of the binary tree as the branch that emerges from the 
root and contains the nodes from which, at each level, the largest number of leaves hang 
(see the bold path in Figure \ref{fig:dendrogram}). The leaves
hanging from each node $x$ of the main branch provide a velocity dispersion $\sigma_{\rm{los}}^x$.
When walking along the main branch from the root to the leaves, $\sigma_{\rm{los}}^x$  
rapidly decreases due to the progressive loss of stars most likely not associated with the dSph (Figure \ref{fig:thresholds}); $\sigma_{\rm{los}}^x$ reaches a plateau when most of the stars hanging from the main branch are members of the dwarf galaxy. In fact, the system is nearly isothermal and progressively removing the less bound stars does not affect the value of $\sigma_{\rm{los}}^x$. Then we arrive at the most bound stars, whose binding energy is very small (as well as their velocity dispersion), so $\sigma_{\rm{los}}^x$ drops again. The extreme of the $\sigma$ plateau closest to the root is thus the appropriate level for the identification of the system.

\begin{figure*}
\center
\includegraphics[angle=0,scale=1.0, bb=30 20 470 280]{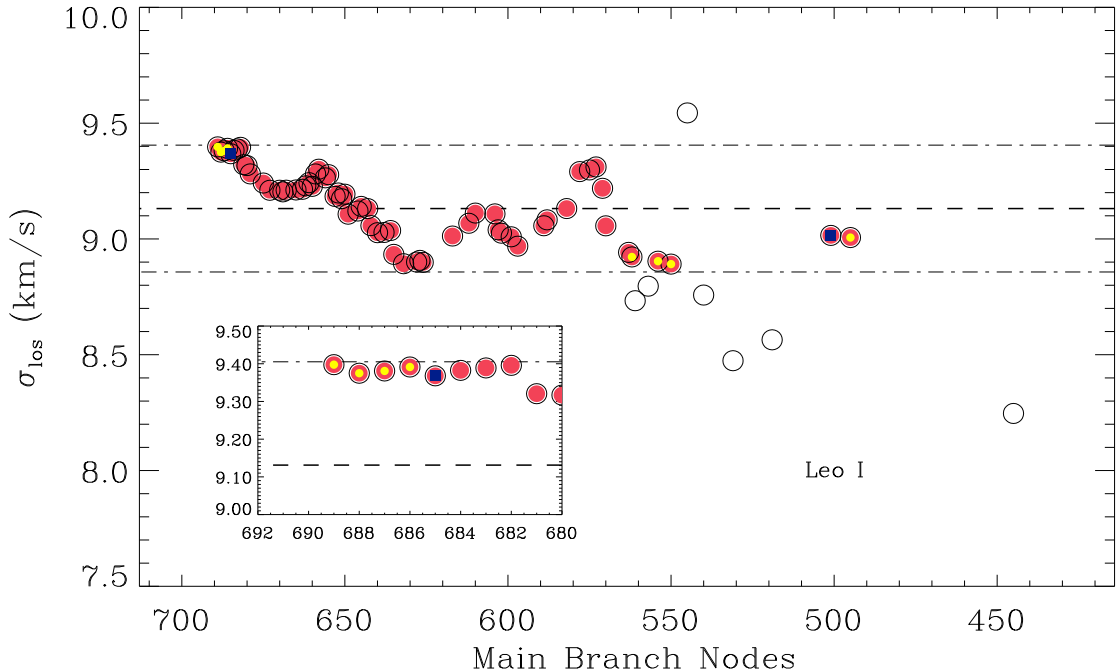}
\vspace{-0.2cm}
\caption{\small{Line-of-sight velocity dispersion of the leaves of each node along the main branch (black circles). The filled red circles are the $N$ nodes. The circles with a yellow center indicate the first and the last 5 nodes of the set $N$ and the blue squares denote the final thresholds.}} 
\label{fig:plateau-def}
\end{figure*}

The procedure to identify the two nodes $x_1$ and $x_2$ corresponding to the $\sigma$-plateau extrema
is as follows (see Figure \ref{fig:plateau-def}). 
Consider the density distribution of 
the population of the $\sigma_{\rm{los}}^x$'s provided by the different nodes $x$.
The $\sigma_{\rm{los}}^x$ that provides the maximum of this density distribution
defines the velocity dispersion
$\sigma_{\rm pl}$, corresponding to the $\sigma$ plateau. We then identify the 
$N$ nodes whose $\sigma_{\rm{los}}^x$ satisfies the relation $\vert 1- \sigma_{\rm{los}}^x/\sigma_{\rm pl}\vert< 
\delta$. The parameters $N$ and $\delta$ clearly depend on each other and 
determine how well we sample the $\sigma$ plateau. To identify the final nodes $x_1$ and $x_2$ from these $N$ nodes, 
we first consider the five nodes closest to the root and the five closest to the leaves;
the node $x_1$ and $x_2$ are the nodes, in the first and second sets respectively, with the minimum deviation from 
$\sigma_{\rm pl}$. With this procedure, we find that $N=15$ provides
the nodes $x_1$ and $x_2$ with $\sigma_{\rm{los}}^x$ sufficiently similar to 
$\sigma_{\rm pl}$ and close enough to the extrema of the $\sigma$ plateau.

The $\sigma$ plateau only appears when the stellar mass $m$ in eq. (\ref{eq:pairwise-energy}) 
lies within a proper range $[m_1, m_2]$, which depends on the dSph considered. Outside this range the plateau is not evident enough and the procedure described in the preceding paragraph does not yield sensible results. 
In fact, each star is a tracer of the velocity and density fields,
namely of the gravitational potential of the system; therefore, 
the parameter $m$ entering eq. (\ref{eq:pairwise-energy}) represents a fraction
of the total mass of the system and is not necessarily close to the real mass of a star. 
The arrangement of the binary tree strongly depends on the parameter $m$ 
because the binding energy (\ref{eq:pairwise-energy}) is a balance
between the potential energy term (proportional to $m^2$) and the kinetic energy term
(proportional to $m$). The absence of a $\sigma$ plateau indicates that 
$m$ assumes an inappropriate value, because the binding energy is dominated 
either by the potential energy or by the kinetic energy. When $m$ is in the correct
range $[m_1,m_2]$, the $\sigma$ plateau appears and we can recognise the bound stellar system.
By varying $m$ within this range, we can estimate the systematic errors on the velocity dispersion
derived from the caustic technique. Below, we apply the caustic
technique for three values of $m$: $m=m_1$, $m=m_2$, and an intermediate value (Table \ref{table3}).

\subsection{Final members}

The stars hanging from the node $x_1$ represent the candidate members of the dSph. They 
determine the centre of the system. The systemic dSph velocity is the median of their velocities, while the celestial coordinates of the centre are the coordinates of the peak
of the two-dimensional density distribution of the stars. 
To find the peak we compute the 2D density distribution on the sky $f_q(\alpha,\delta)$ with the adaptive kernel method described in \cite{diaf99}. The centres we determine coincide, within the uncertainties, with those provided in the literature.

\begin{figure}
\center
\includegraphics[angle=0,scale=0.6,bb=35 14 453 305]{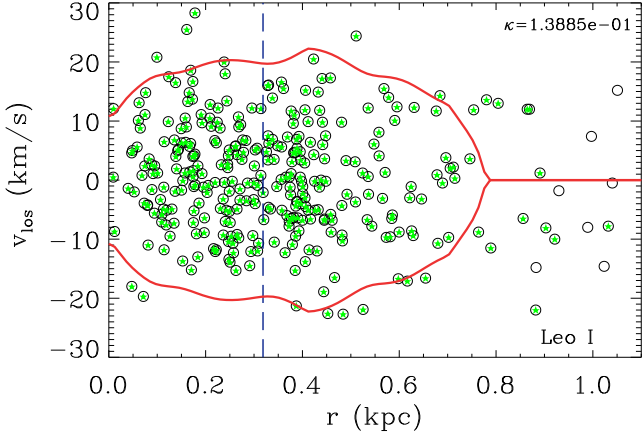}
\vspace{-0.2cm}
\caption{\small{Velocity diagram for Leo I (data from \citealt{mateo07}). The black circles are all the stars in the catalogue, while the circles with a green star are the candidates selected by the binary tree. The red curves are the caustics identifying the members. The blue vertical line represents the mean separation $R$ from the centre.}} 
\label{fig:red-diag}
\end{figure}

In the plane $(r,v)$ (Figure \ref{fig:red-diag}), where $r$ is the projected distance and $v$ the los 
velocity of each star from the dSph centre, the caustics are the curves satisfying the equation 
$f_q(r,v)=\kappa$. Here $f_q(r,v)$ is the star distribution in the plane, and $\kappa$ is the root of the equation
\begin{center}
\begin{equation}
\langle v_{\rm esc}^2\rangle_{\kappa,R}=4\langle v^2\rangle\; .
\label{eq:skappa}
\end{equation}
\end{center}
The function $\langle v_{\rm esc}^2\rangle_{\kappa,R}=\int_0^R{\cal A}_\kappa^2(r)\varphi(r)dr/
\int_0^R\varphi(r)dr$ is the mean caustic amplitude within $R$,
$\varphi(r)=\int f_q(r,v) dv$, $\langle v^2\rangle^{1/2}$ is
the velocity dispersion of the candidate members 
and $R$ is their mean projected separation from the centre. 
The stars within the upper and lower caustics are the dSph members, used to compute $\sigma_{\rm los}(r)$.

We can use this set of members as the input of an iterative procedure:
we locate the caustics in the velocity diagram, remove the stars outside the caustic, and
locate new caustics with only the stars left. We proceed until no star is rejected.
For each dwarf, we apply this procedure to the velocity diagram using the outcomes 
provided by the binary tree corresponding to each stellar mass: the
procedure converges after a number of iterations in the range $[3,13]$, 
and the total number of rejected stars 
is in the range  $[3,45]$, while the percentage of the rejected stars is $[0.2,5.7]\%$ of the stars within the initial caustics.
We conclude that the iterative procedure adds little to the 
identification of members: both the initial
and final caustics and the initial and final velocity dispersion profiles are basically indistinguishable.
Nevertheless, we keep the set of stars provided by the iterative procedure as the final members.

\begin{figure*}
\center
\includegraphics[angle=0,scale=0.7,bb=0 0 566 255]{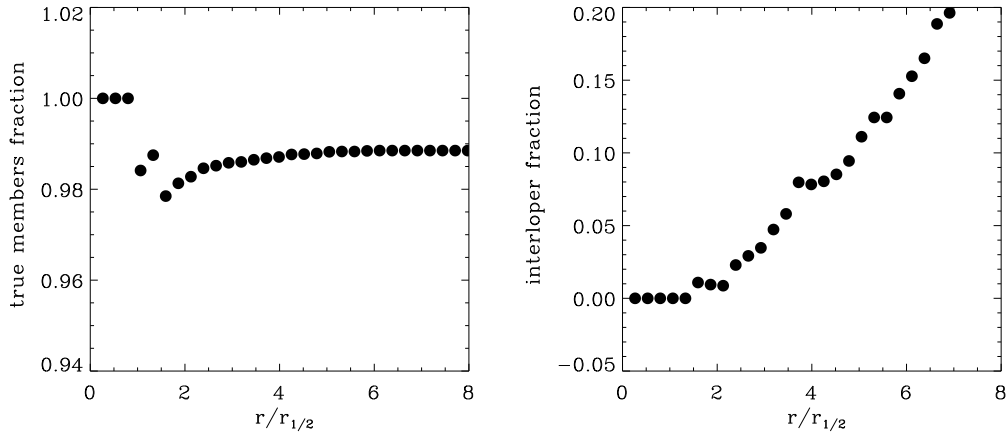}
\vspace{-0.2cm}
\caption{\small{Identified members and interloper fraction resulting from the application of the caustic technique to a simulated halo.}} 
\label{fig:members}
\end{figure*}

We performed extensive tests of the caustic technique as an 
identifier of interlopers of self-gravitating systems. We used dark 
matter halos extracted from N-body 
simulations of a cubic volume of $192$~$h^{-1}$~Mpc on a side of 
a standard $\Lambda$CDM model \citep{borgani04}. We will present our results elsewhere (Serra et al. in preparation).
Here, we provide in advance an example of our tests, to show that the caustic 
technique is indeed rather effective at identifying interlopers in simulated dark matter halos. 
For the sake of simplicity, we define the members as the particles within the sphere of 
radius 3$r_{200}$, where $r_{200}$ is the radius within which the average mass density
is 200 times the critical density $3H_0^2/8\pi G$. 
Figure \ref{fig:members} shows the cumulative member fraction at increasing
radii $r$ of a dark matter halo from our simulation: 
$r$ is in units of the half-mass radius $r_{1/2}$, defined by the relation
$M(<r_{1/2})=M(<r_{200})/2$. For this halo, $r_{200}=2.25r_{1/2}$. The caustic 
technique identifies more than 98\% of the real members at all radii (left panel). Moreover,
the fraction of interlopers, i.e. the particles misidentified as members, remains below 10\% out
to $5r_{1/2}$ and increases to 20\% only at radii as large as $7r_{1/2}$ (right panel).

\begin{figure*}
\center
\includegraphics[angle=0,scale=0.85,bb=20 14 623 297]{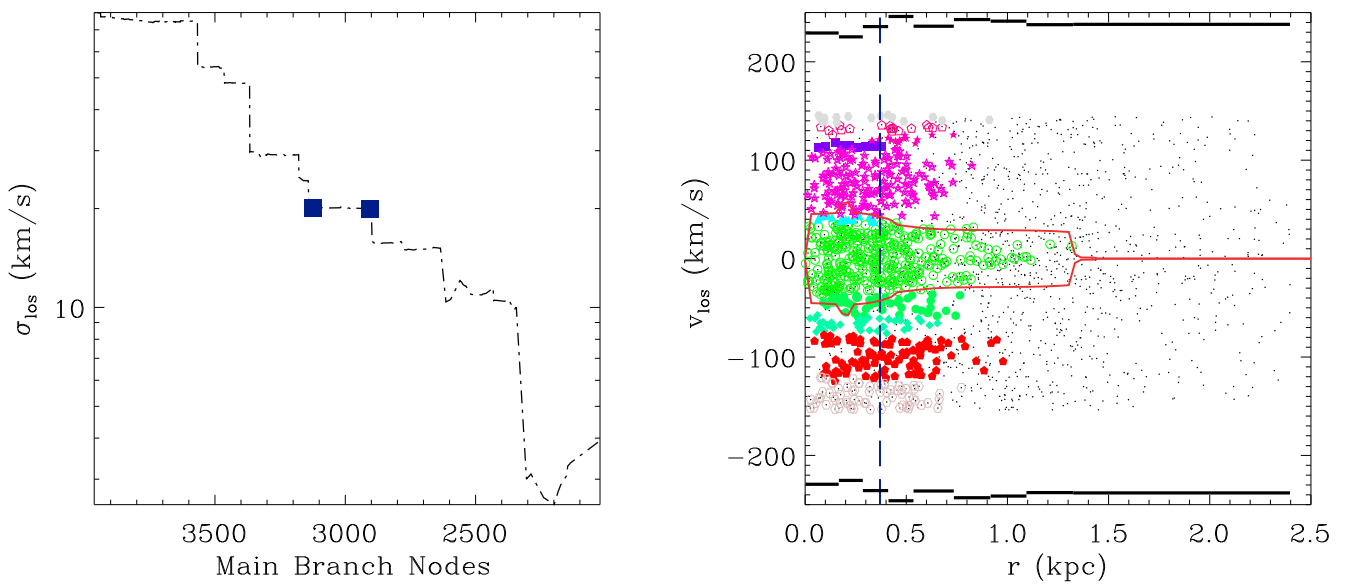}
\vspace{-0.2cm}
\caption{\small{Velocity dispersion along the main branch and redshift diagram of a mock dSph in a dense environment. The different colours represent the groups identified by the binary tree (the green circles are the stars of the main group). The thick solid lines delimit the 3$\sigma$ range.}} 
\label{fig:3sigma}
\end{figure*}

We emphasise the
robustness of the caustic technique by comparing it to the 3$\sigma$-clipping method. Figure \ref{fig:3sigma} shows a dSph embedded in a particularly dense environment. We have built this catalogue by taking the celestial coordinates of the stars in the field of Carina and assigning a los velocity to each star in such a way to create a uniform velocity distribution across the range $[-30,30]$~km~s$^{-1}$ and a uniform background in the velocity space along $[-150,150]$~km~s$^{-1}$. The membership determined
by the 3$\sigma$-clipping procedure 
yields a velocity dispersion of $\sim 80$~km~s$^{-1}$, a
factor of four larger than the real velocity dispersion of the main distribution $\sim 17$~km~s$^{-1}$, while the caustic technique gives $\sim 20$~kms$^{-1}$.  
The worse performance of the
3$\sigma$-clipping procedure is a consequence of its assumption of an underlying Gaussian distribution 
of the line-of-sight velocities. This assumption is unnecessary in the caustic
technique, which therefore appears to be less prone to failure when the foreground and background
contamination is severe. We have chosen to perform this test, even if this system is unrealistic, to show that the technique yields reliable results in a dense velocity diagram.

The caustic technique was already 
used to remove interlopers both in real clusters \citep[e.g.][]{Rines04, Rines05},
to investigate the dependence of galaxy properties on environment, 
and the stellar halo of the Milky Way \citep{brown09}, to estimate
the velocity dispersion of the stars in the halo.

\section{Results}

\begin{figure*}
\centering
\includegraphics[angle=0,scale=.55,bb=20 14 453 325]{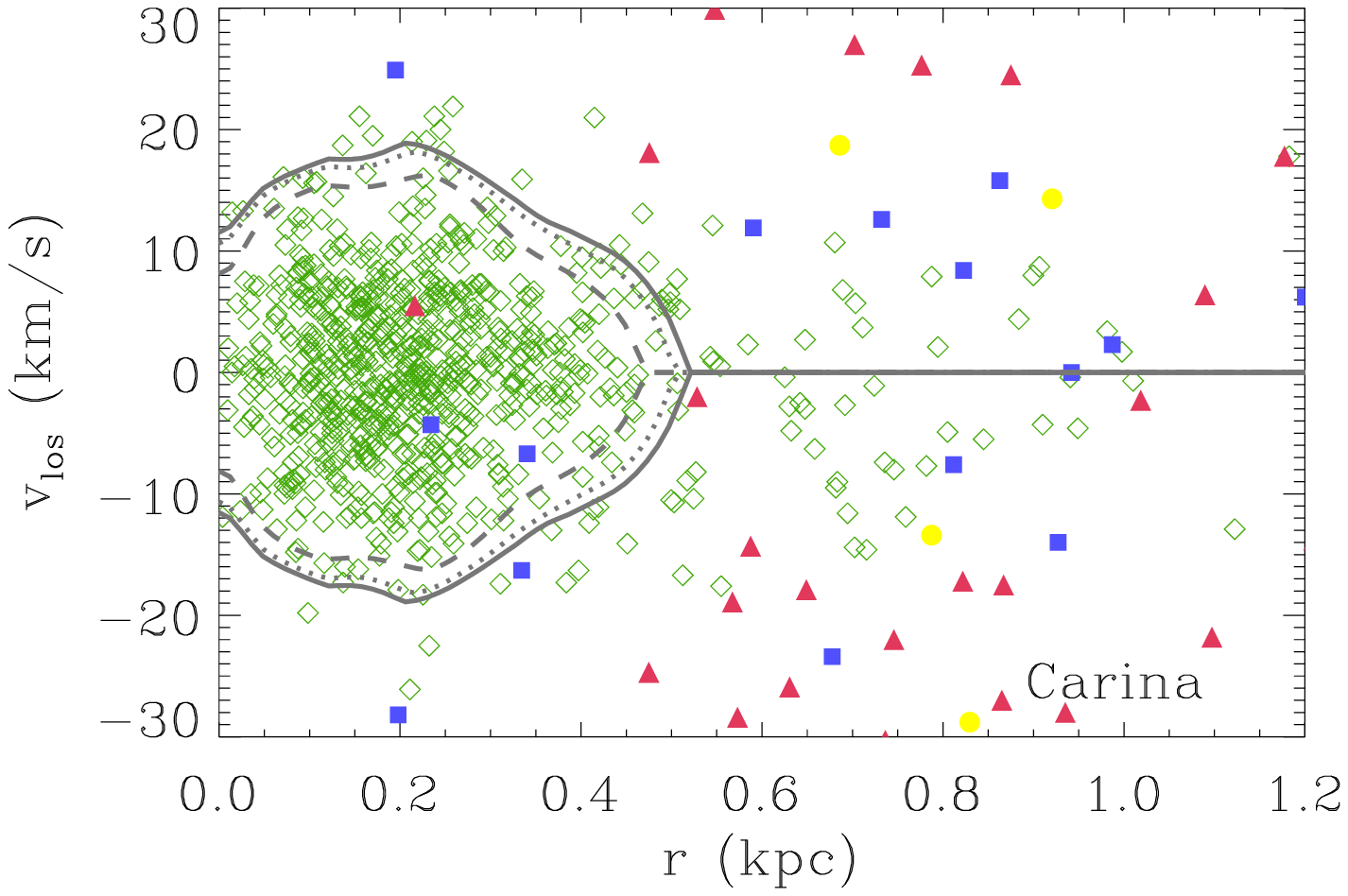}
\includegraphics[angle=0,scale=.55,bb=20 14 453 325]{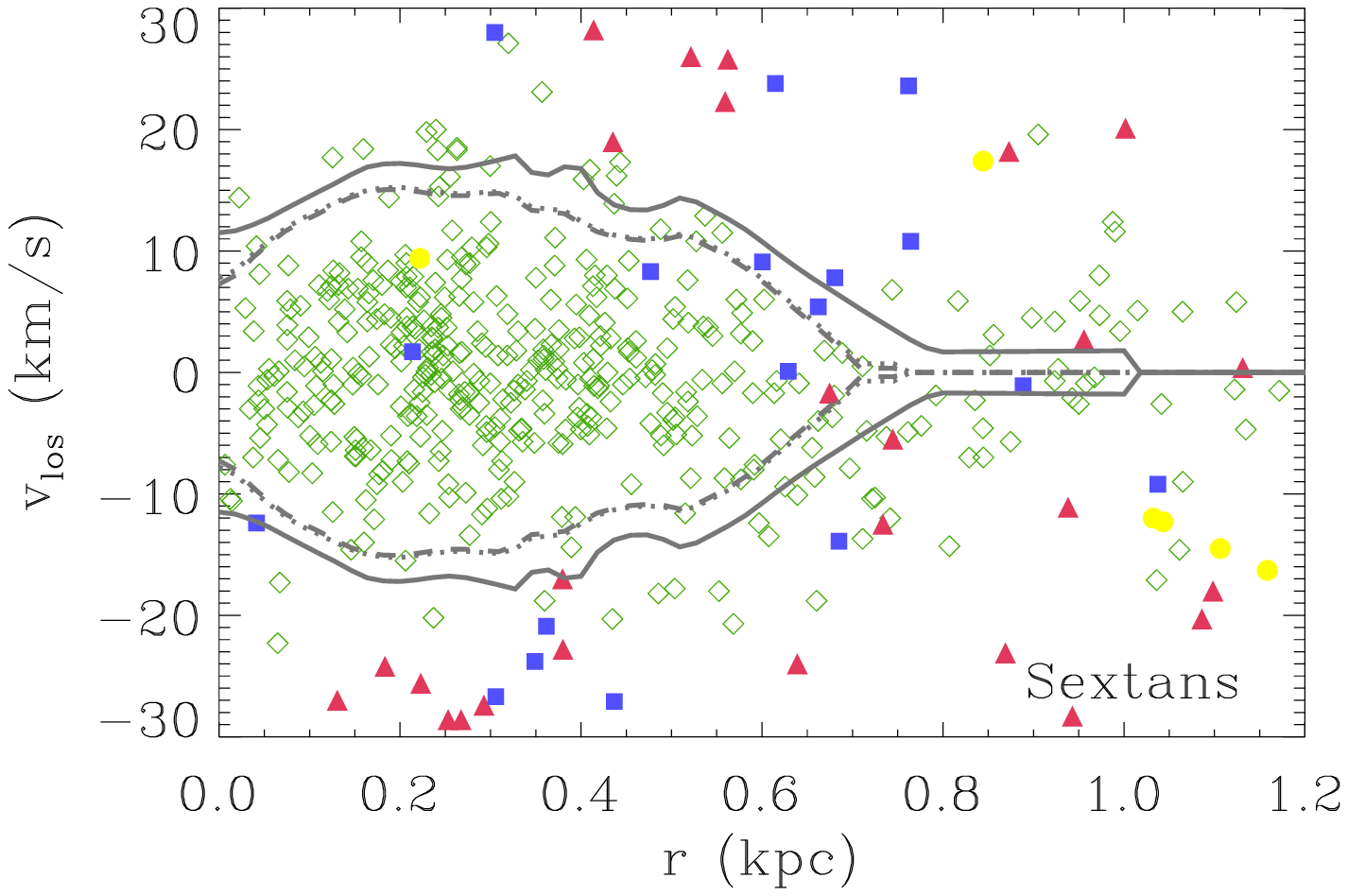}
\includegraphics[angle=0,scale=.55,bb=20 14 453 325]{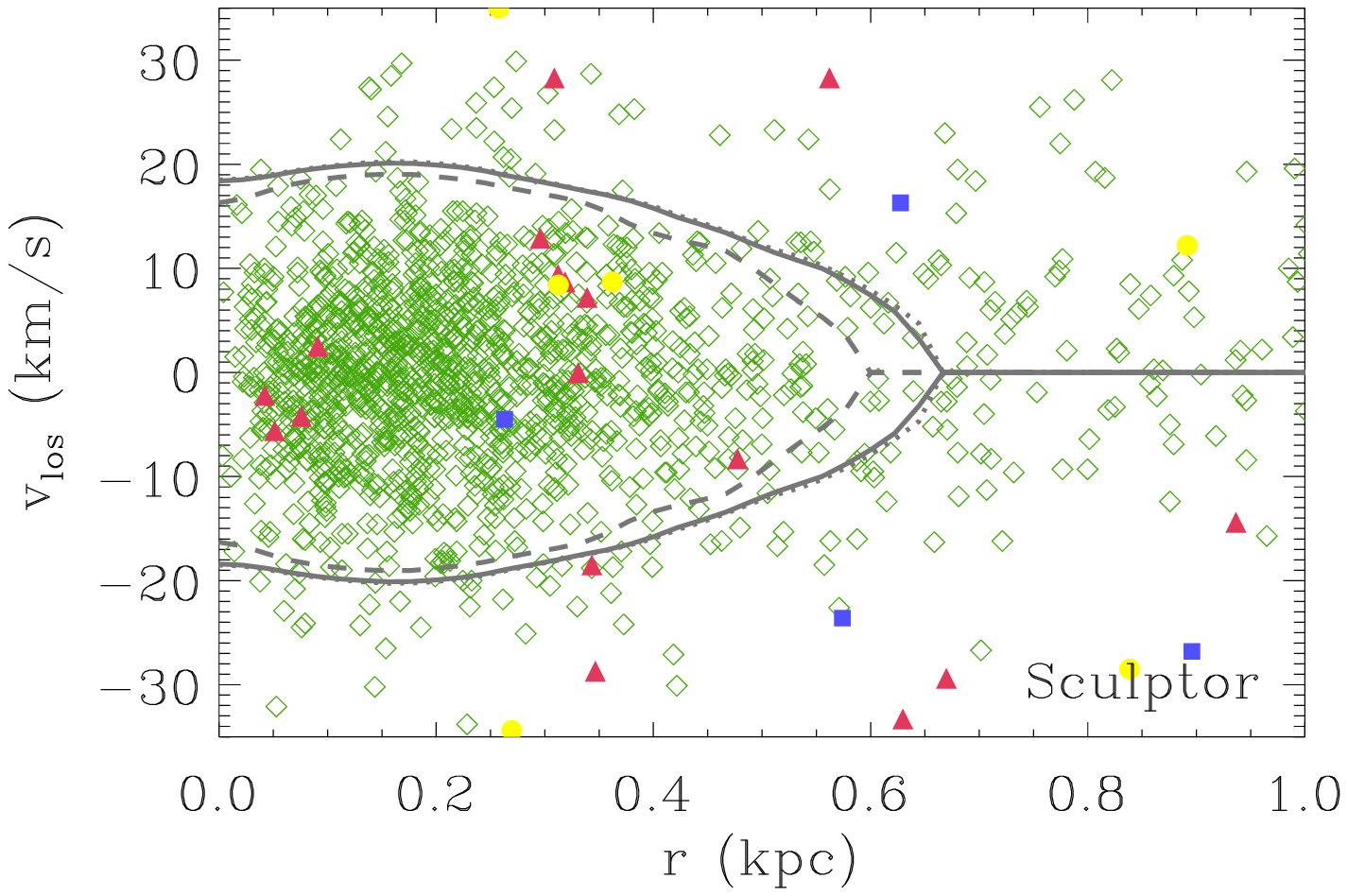}
\includegraphics[angle=0,scale=.55,bb=20 14 453 325]{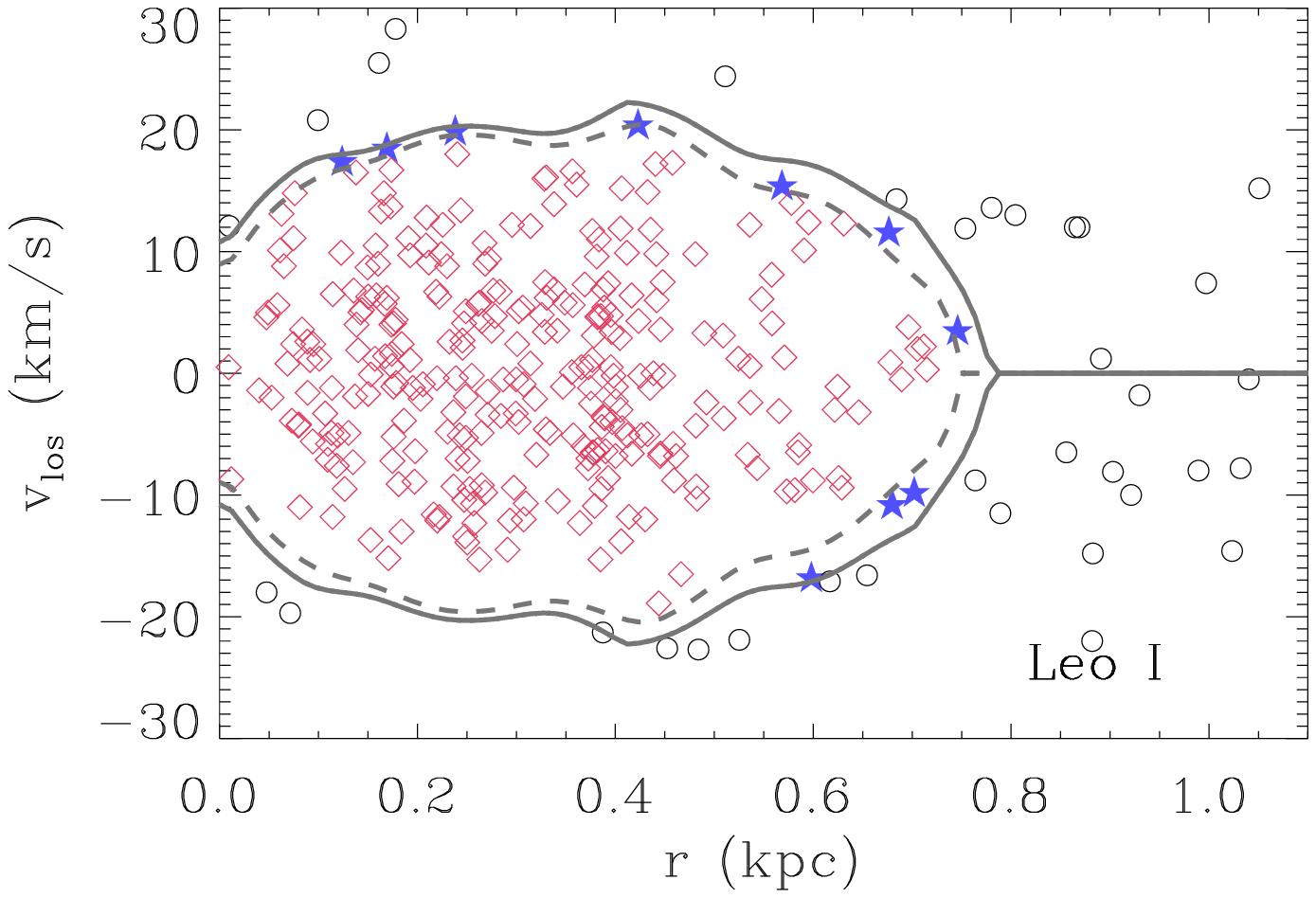}
\includegraphics[angle=0,scale=.55,bb=20 14 453 325]{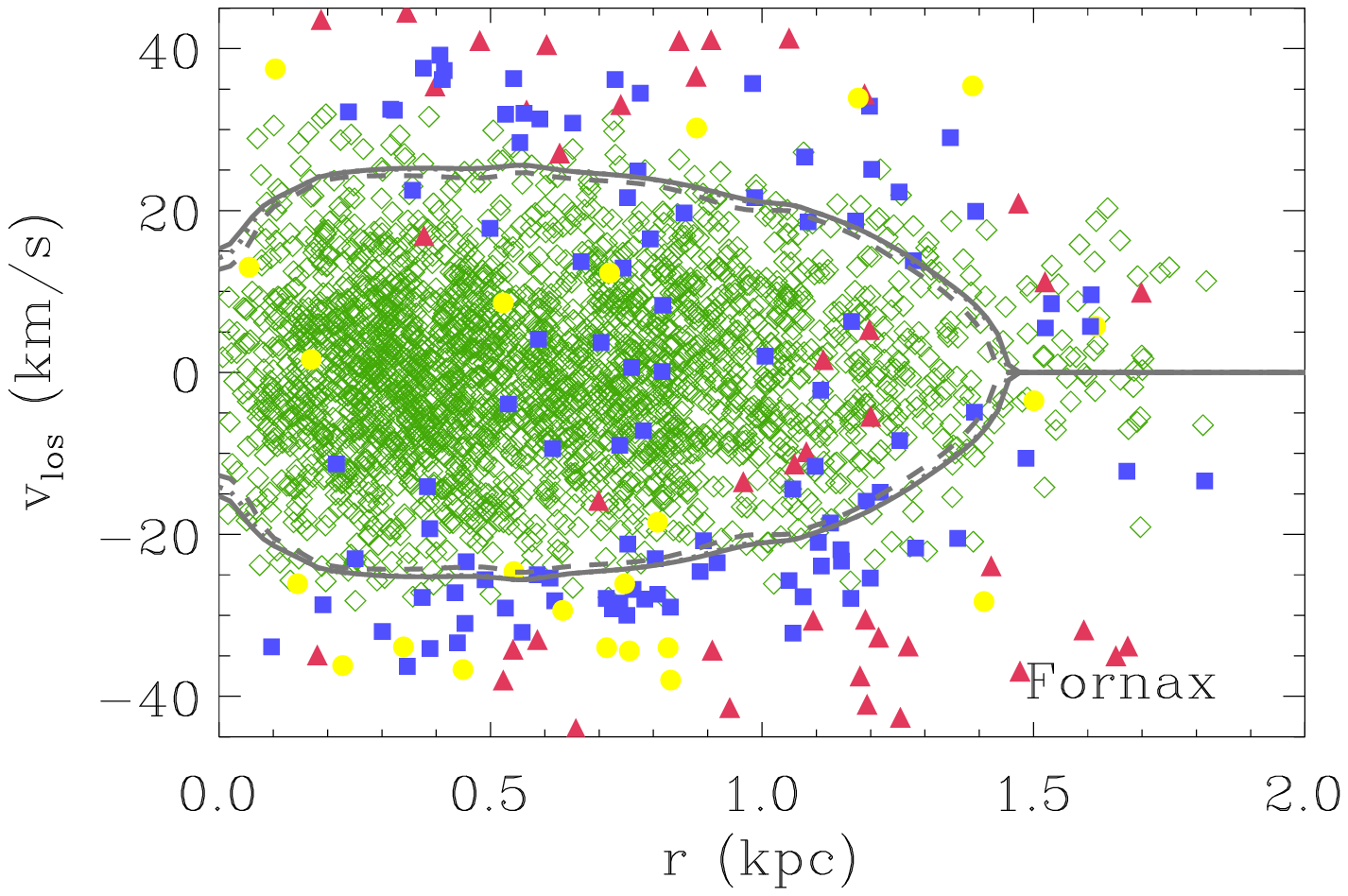}
\caption{\small{Velocity diagrams for our 5 dSphs with the 3 sets of caustics overplotted (see \S\ref{sec:starmass}). For Leo I: the dark pink diamonds indicate stars that are members of all three caustics, the black circles are stars that are members of no caustics and the blue stars are members of at least one, but not all the caustics. For the remaining dwarfs: the three set of caustics are the dotted, dashed and solid lines, the colours of the symbols indicate the probability $p$ of being members. $p \leqslant 0.5$: red triangles, $0.5 \leqslant p \leqslant 0.7$: yellow circles, $0.7 \leqslant p \leqslant 0.9$: blue squares, $0.9 \leqslant p$: green diamonds according to Walker et al. (2007).}}
\label{fig:caustics}
\end{figure*}

\begin{figure*}
\centering
\makebox[12cm][c]{\hspace{1.5cm}\textbf{Carina}}
\includegraphics[angle=0,scale=.48,bb=50 40 506 277]{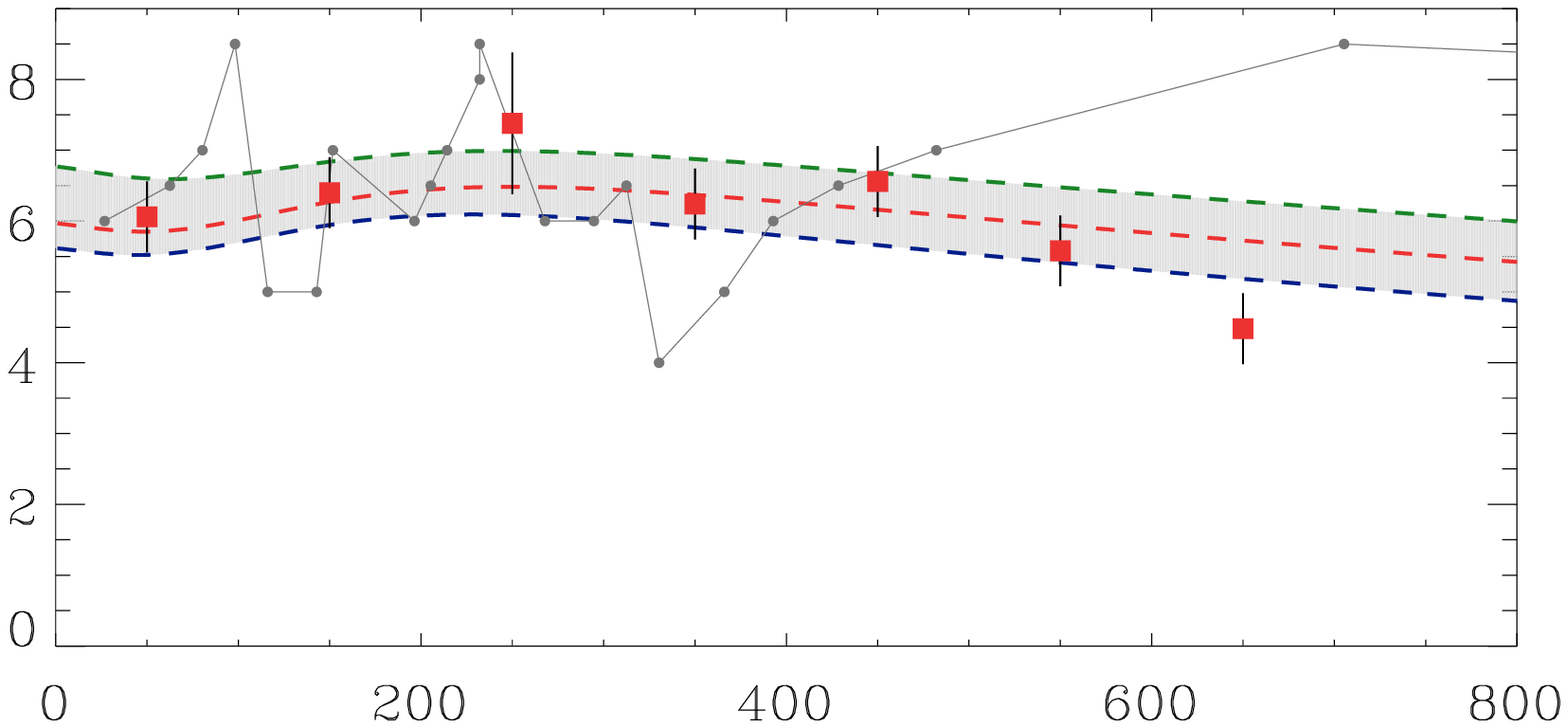}
\includegraphics[angle=0,scale=.48,bb=20 40 506 277]{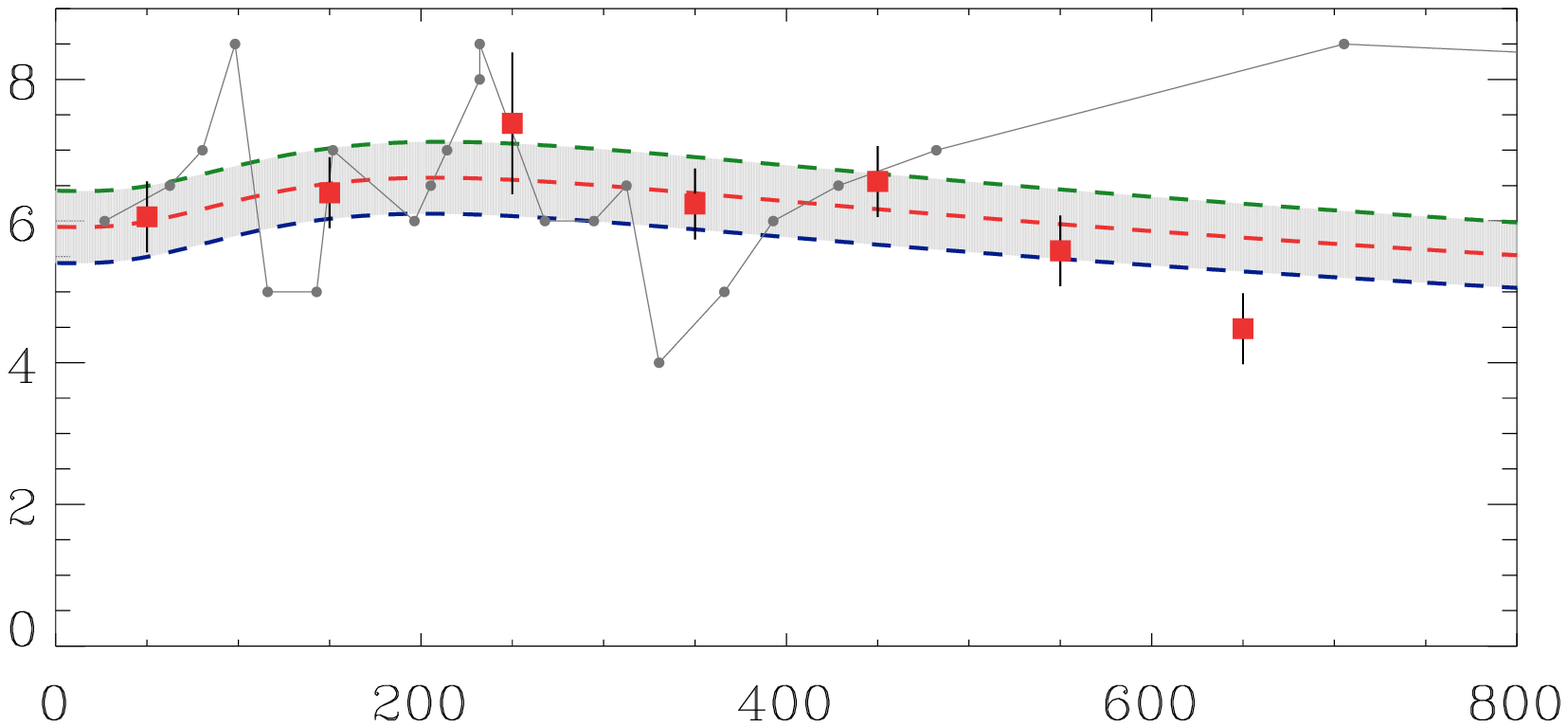}
\makebox[12cm][c]{\hspace{1.5cm}\textbf{Sextans}}
\includegraphics[angle=0,scale=.48,bb=50 40 506 27]{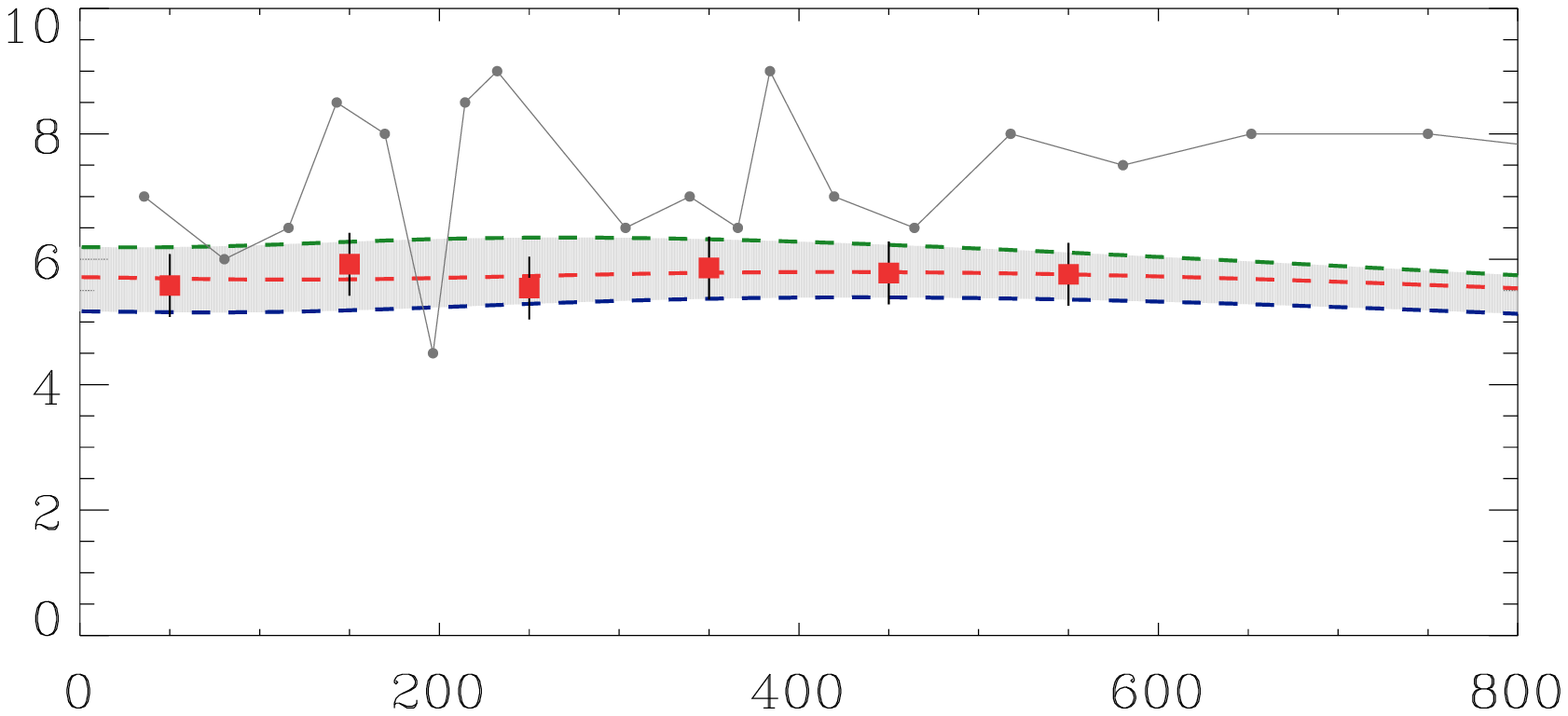}
\includegraphics[angle=0,scale=.48,bb=20 40 506 277]{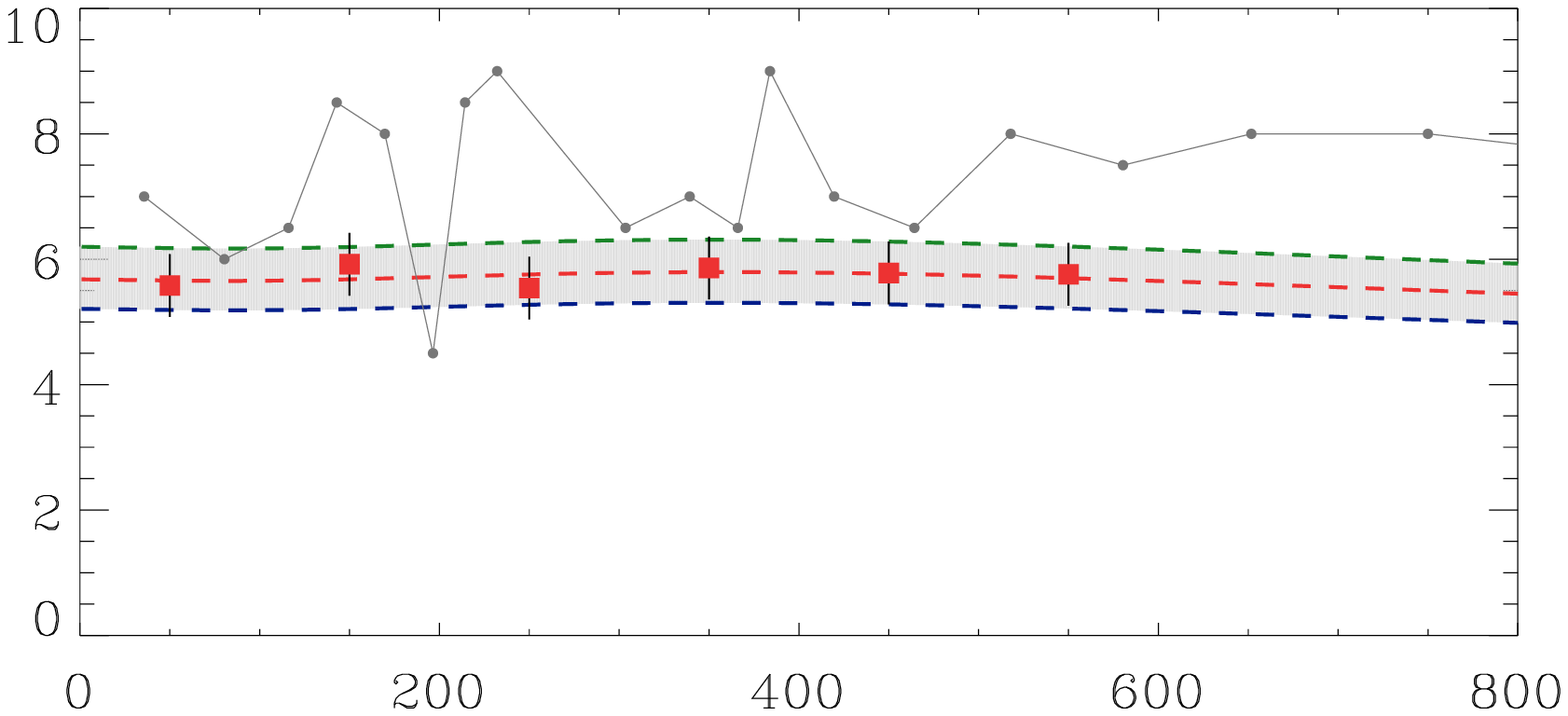}
\makebox[12cm][c]{\hspace{1.5cm}\textbf{Sculptor}}
\includegraphics[angle=0,scale=.48,bb=50 40 506 277]{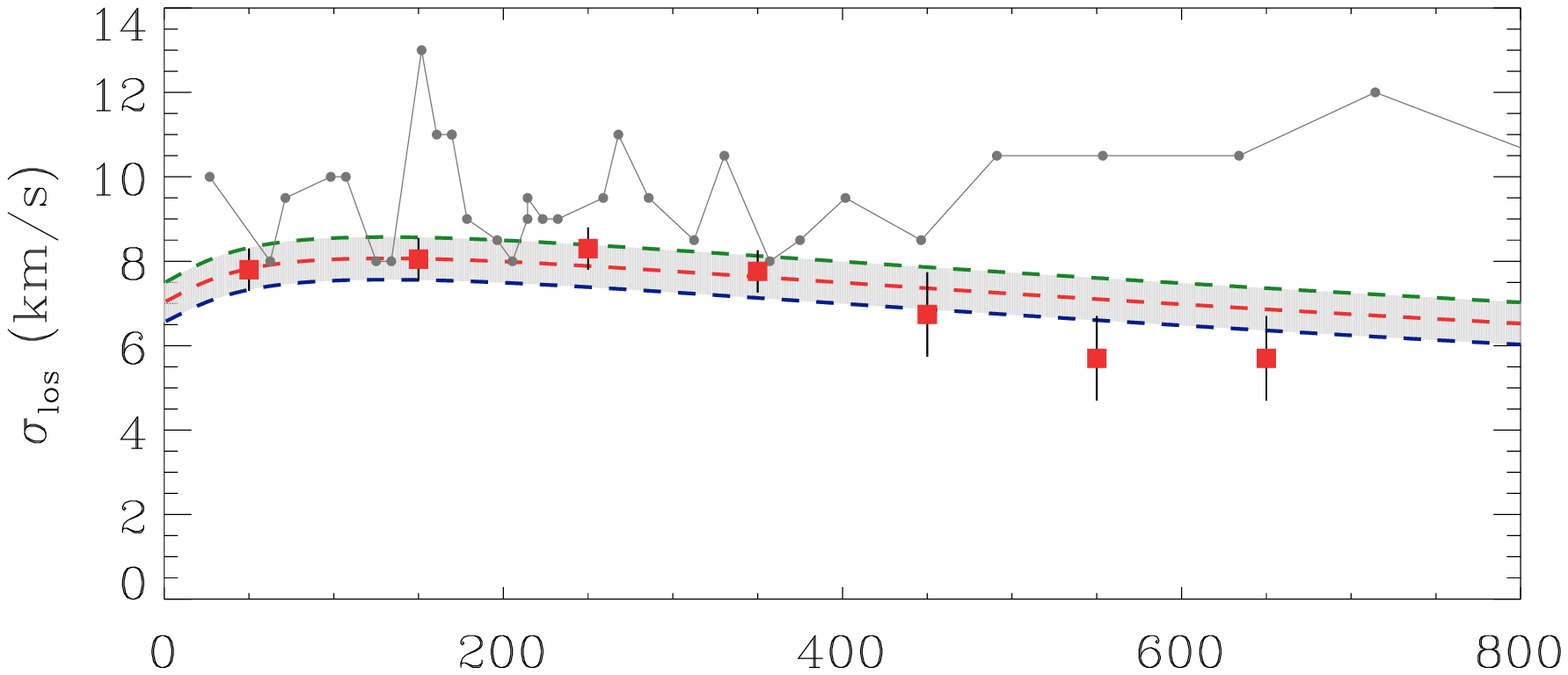}
\includegraphics[angle=0,scale=.48,bb=20 40 506 277]{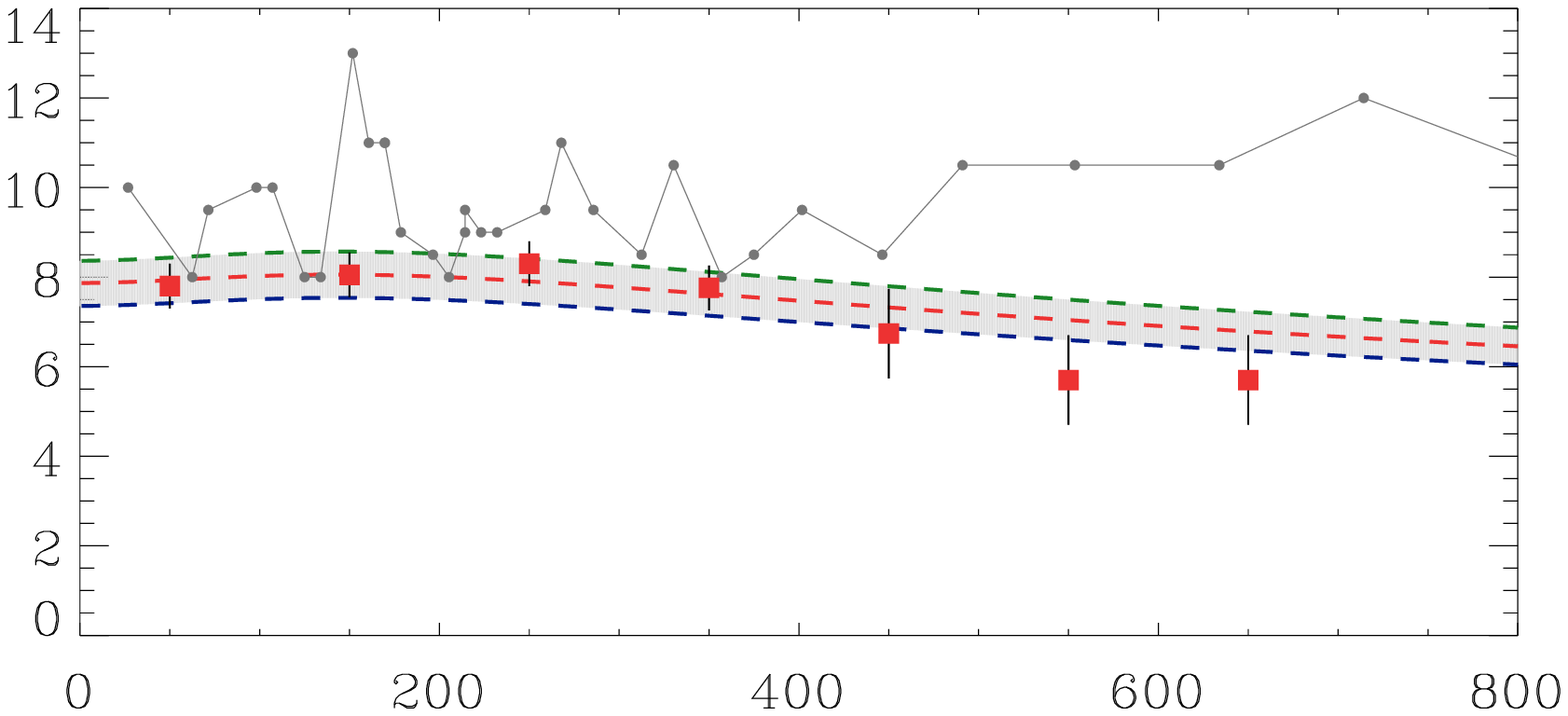}
\makebox[12cm][c]{\hspace{1.5cm}\textbf{Leo I}}
\includegraphics[angle=0,scale=.48,bb=50 40 506 277]{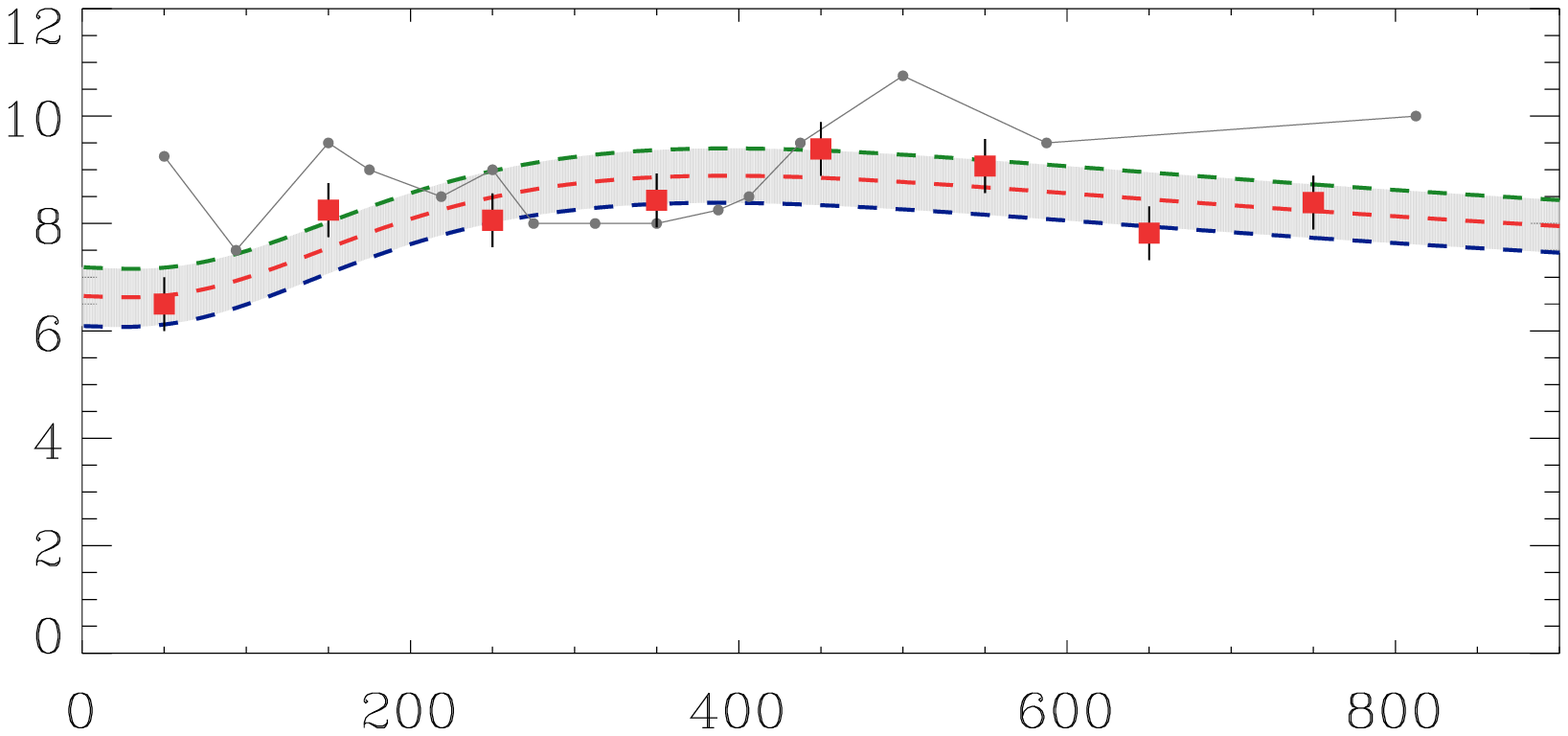}
\includegraphics[angle=0,scale=.48,bb=20 40 506 277]{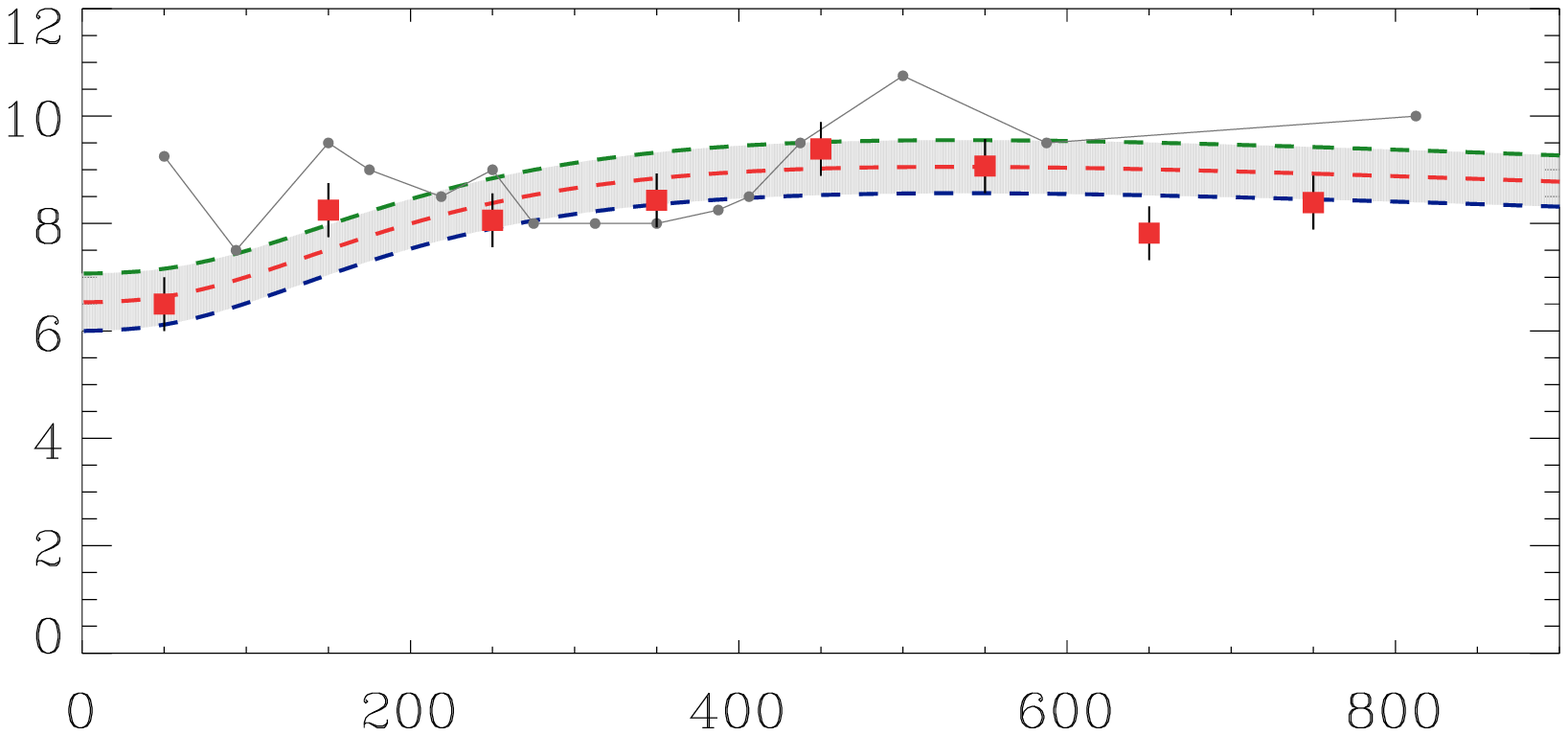}
\makebox[12cm][c]{\hspace{1.5cm}\textbf{Fornax}}
\includegraphics[angle=0,scale=.48,bb=50 40 506 277]{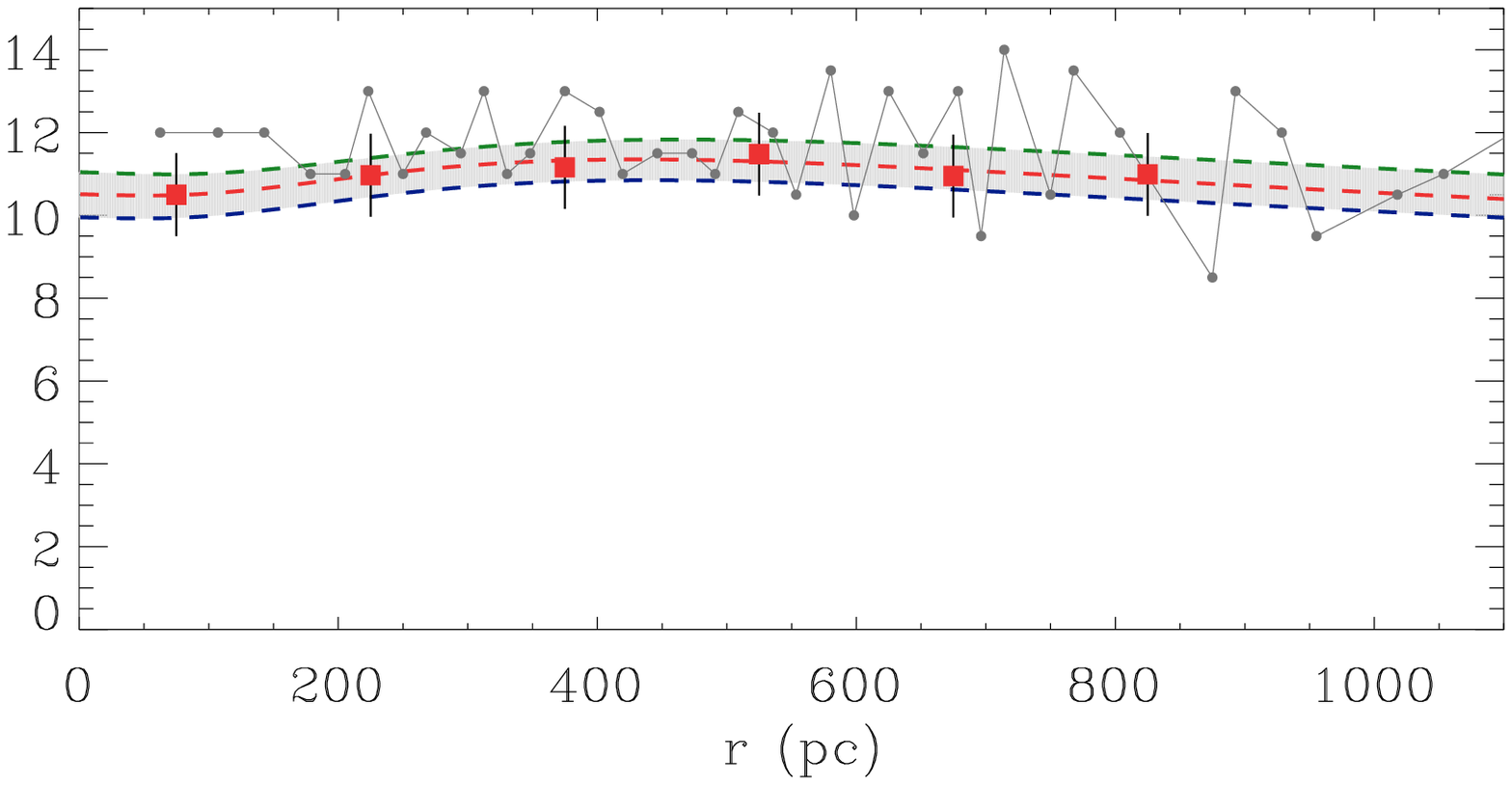}
\includegraphics[angle=0,scale=.48,bb=20 40 506 277]{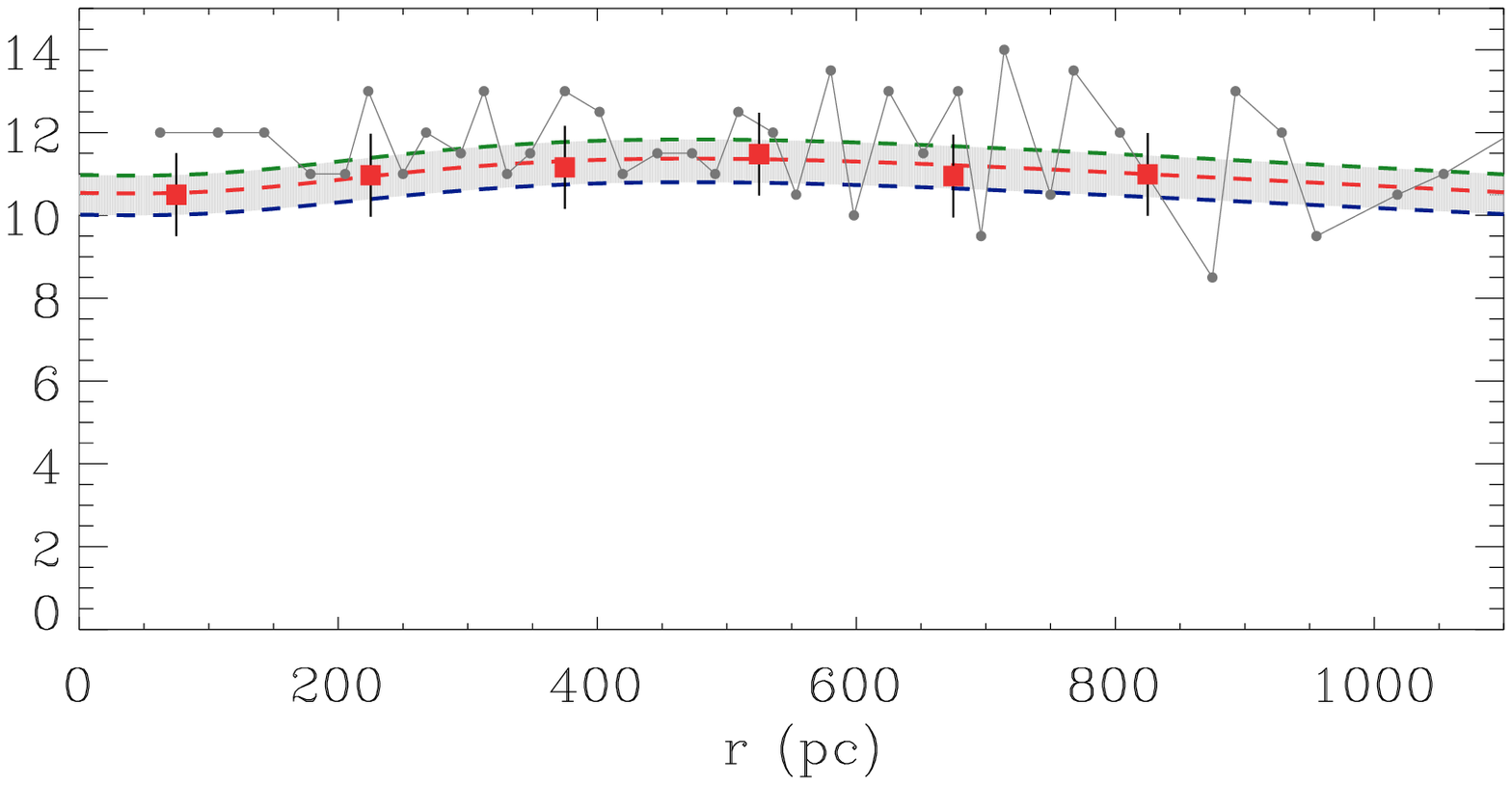}
\caption{\small{Line-of-sight velocity dispersion profiles for the Walker et al. (2007) uncleaned samples (grey circles) and our cleaned samples (red squares). In the left-hand panels we overplot the best fit MOND models (parameters listed in Table 2) and in the right-hand panels we plot the best fit cored dark matter halo models (parameters in Table 1). The two lines either side of the best fit are the 1-$\sigma$ errors on the $\rho_{DM,0}$ and $r_{DM}$ parameters. }} 
\label{fig:veldisp}
\end{figure*}

For the five dSphs, we separate the member stars identified by the caustic technique into projected radius bins of 50~pc. 
We then evaluate the standard deviation of the velocities around the systemic velocity, which gives us the los velocity dispersion (losVD) as a function of the projected radius. 
Figure \ref{fig:caustics} shows the velocity diagram of the five dSphs and Figure \ref{fig:veldisp} shows the cleaned losVDs for the star members
identified with the intermediate of the three sets of caustics that we locate for the three adopted masses (Table \ref{table3}). Given that we use the entire 
sample of stars, in Figure \ref{fig:caustics} we plot the probability of membership for each star, according to \cite{walker09}. In all cases there are a number of low-probability stars within 
the caustics; however, if we exclude these stars, the velocity dispersion profiles remain unaffected. In Figure \ref{fig:veldisp}, the losVD as calculated by \cite{walker07} are also overplotted for reference. 

The losVD computed through the caustic technique are considerably smoother and in some cases lower than those calculated by \citet{walker07}. Moreover, the number of members identified with the caustic technique
in each dwarf is always smaller than the number of members identified by the algorithm of \citet{wal09b} 
based on the joint distribution of velocity and the magnesium strength. Our
procedure of interloper removal also appears to be more strict than the technique of \citet{klimentowski08}.
According to Figure 1 of \citet{lok09}, who use \citet{klimentowski08}'s procedure, 
the number of members she finds is 5-10\% greater than our results for Carina, Fornax and Sculptor; for Sextans, her 
number of members is ~25\% larger.

The caustic technique has not significantly altered the
magnitude of the losVD for the most distant and luminous dSphs (Leo I and
Fornax): for these larger dSphs, it has merely pruned some outlying stars.
However, for Sextans and Sculptor, the losVDs have changed considerably.

The typical error bars on each losVD bin in \cite{walker07} are $\pm1.5~\kms$, but we have used fewer bins, thus increasing the number of stars per bin and decreasing the uncertainty
on each losVD value. Additionally, the three sets of losVDs identified by the caustic technique, found using different star masses, do not systematically differ by more than $\pm 0.5~\kms$ until the last data point for Carina, Sextans and Sculptor (see Figure~\ref{fig:veldisp_caus}). The three caustics of Fornax and Leo I vary even less. Therefore, we take $\pm$0.75~$\kms$ as a reasonable error bar for each of our radius bins.

For each of the five dSphs, we performed a Jeans analysis to infer their DM halo
properties or to infer their $M/L$ ratios in the case of MOND. Crucially, the Jeans equation is
\begin{equation}
\label{eqn:jeans}
{d \over dr}\sigma_r^{2}(r) + \left[\alpha(r)+2\beta(r) \right]r^{-1}\sigma_r^{2}(r) = -g(r)
\end{equation}
which solves for the radial velocity dispersion, $\sigma_r^{2}(r)$, making
use of $\alpha(r)={d\ln \rho_* \over d\ln r}$. From this we use the
standard method to cast the radial velocity dispersions into los
velocity dispersions. The vital point for our analysis is that $g(r)$ is the
only part of the equation to change between the MOND and dark matter
scenarios. Using dark matter, we know $g(r)=g_*(r)+g_{DM}(r)$. We discuss
the case of MOND in \S\ref{sec:mond}.

All the dSph parameters, such as luminosity and distance, are taken from the table of \cite{angus08} - where the main observational references are \cite{mateo98} and \cite{irwinhatz}.

\subsection{Dark halo properties}

To demonstrate how quantitatively different the DM
halos required to fit the cleaned losVDs are, we used the cored dark halo density profile $\rho(r)=\rho_{DM,0}[1+(r/r_{DM})^2]^{-1.5}$ (cusped halos can achieve just as good fits as shown in \citet{angusdiaf09} and discussed in \citealt{eva09}) 
applied in \cite{angusdm} (where we refer interested readers to find a detailed description of the models) along with constant,
velocity anisotropies ($\beta(r)=\beta_o$). We tried models with radially varying velocity anisotropy, but found no improvement in the fits.

We began by evaluating the $\chi^2$ on a finely spaced grid of the free parameters: $\rho_{DM,0}$ and $r_{DM}$ in the dark halo scenario, and $\rho_{*,0}$ (or $M/L$) in the case of MOND, as well as velocity anisotropy in both cases. The unreduced $\chi^2$ is defined by
\begin{equation}
 \chi^{2}=\sum_{i=1}^{n}\left[\frac{(\sigma_{los}(r_i)-\sigma_{obs}(r_i))^{2}}{err^2(\sigma_{obs}(r_i))}\right]
\end{equation}
where $r_i$ is the radius of the $i^{th}$ data point, $\sigma_{los}$ is the fit and $\sigma_{obs}$ is the observed velocity dispersion along the line of sight.

After finding the $\chi^2$ minimum for each dSph, we re-evaluated the surrounding parameter space with greater resolution, and to determine the 1-$\sigma$ error on each parameter. This is given by the difference from the best fit value a parameter must be altered in order to increase the $\chi^2$ by 3.5 in the case of the dark halos with three free parameters. Figure \ref{fig:forr} demonstrate how in the case of MOND, the $\chi^2$ depends on the two free parameters $M/L$ and velocity anisotropy and from this the 1-$\sigma$ errors are easily deduced.

Apart from Fornax, all the other dSphs are consistent at 1-$\sigma$ with any $\beta$ between -3 and -0.5, with a suitably chosen  $\rho_{DM,0}$ and $r_{DM}$. For higher $\beta$ (i.e. greater than -0.5) there is always a point where for any combination of $\rho_{DM,0}$ and $r_{DM}$ the $\Delta\chi^2$ above the $\chi^2$ minimum reaches 3.5 (or 1-$\sigma$ for 3 free parameters). On the other hand, for very negative $\beta$, it is not always possible to reach a $\Delta\chi^2$ of 3.5. However, only in one case (Carina) does the best fit prefer lower values of $\beta$ than -3, and even then the difference in $\chi^2$ between $\beta=-1.3$ and -3 is negligible (see Figure \ref{fig:carb}). For this reason, in Table 1 the lower (upper) error bar on $\rho_{DM,0}$ ($r_{DM}$) is not strictly the 1-$\sigma$ error, but that value at $\beta=-3$ (and the noted $\chi^2$). Here we are applying a prior that the velocity anisotropy cannot be highly tangentially biased.

The parameters of all the models are given in Table 1 with their 1-$\sigma$ error. Clearly there is a large range of central DM densities ($\rho_{DM,0}$) and DM scale radii ($r_{DM}$) which can achieve a good fit to the data and this range is only truncated for large  $\rho_{DM,0}$ and low $r_{DM}$ because we force the condition $\beta > -3$.

\begin{figure*}
\centering
\includegraphics[angle=0,scale=0.45,bb=20 14 651 439]{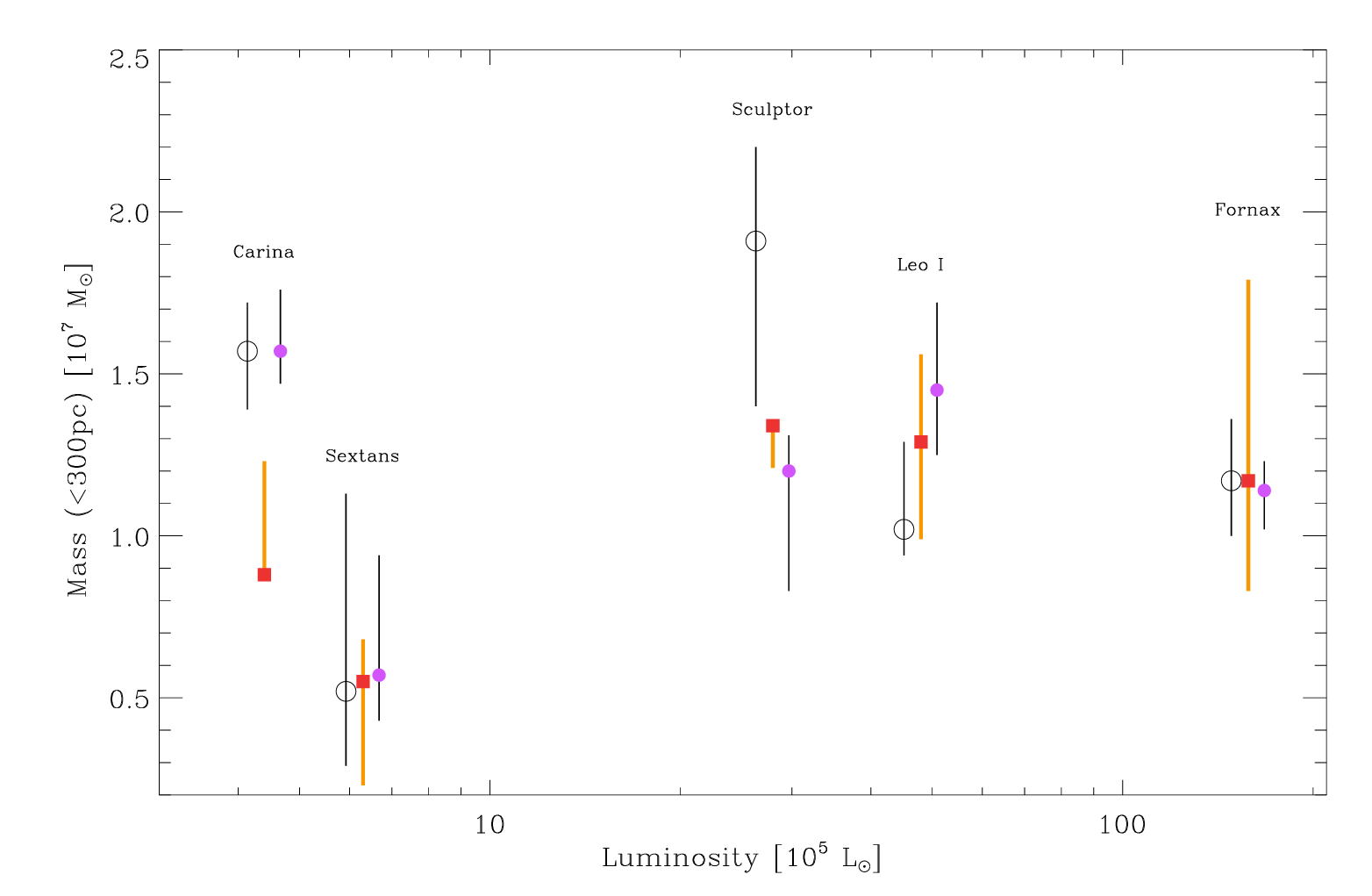}
\vspace{-0.2cm}
\caption{\small{The enclosed dark mass within 300~pc as a function of dSph luminosity. The purple dots are from Strigari et al. (2008) and the red squares are our cleaned samples presented in Figure \ref{fig:veldisp}. The empty circles correspond to the fits of Angus (2010) to the uncleaned Walker et al. (2007) data, using the same cored dark halos as we implement here. The points referring to the same dSph are slightly shifted along the L axis for clarity.}}
\label{fig:ML}
\end{figure*}

\begin{figure}
\centering
\includegraphics[angle=0,scale=0.5,bb=80 360 550 720]{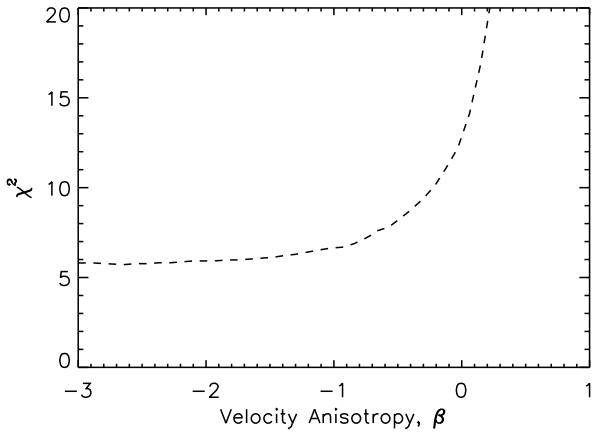}
\vspace{-0.2cm}
\caption{\small{The lowest possible $\chi^2$ for Carina with a fitted dark halo as a function of velocity anisotropy. Notice the lowest $\chi^2$ decreases only very slightly from $\beta$=-1.3 to -3.}}
\label{fig:carb}
\end{figure}

\begin{figure*}
\centering
\includegraphics[angle=0,scale=0.5,bb=80 360 550 720]{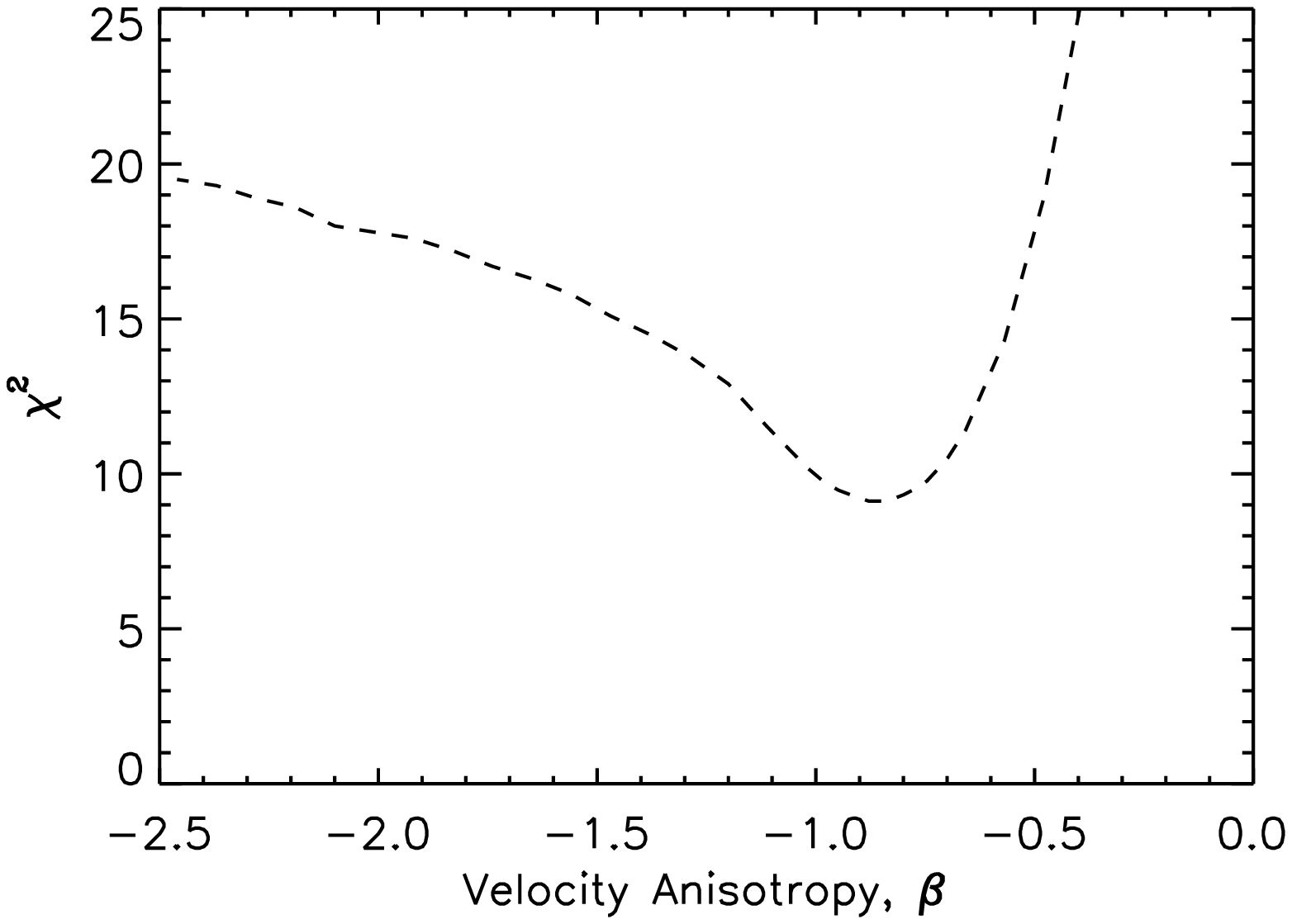}
\includegraphics[angle=0,scale=0.5,bb=60 360 550 720]{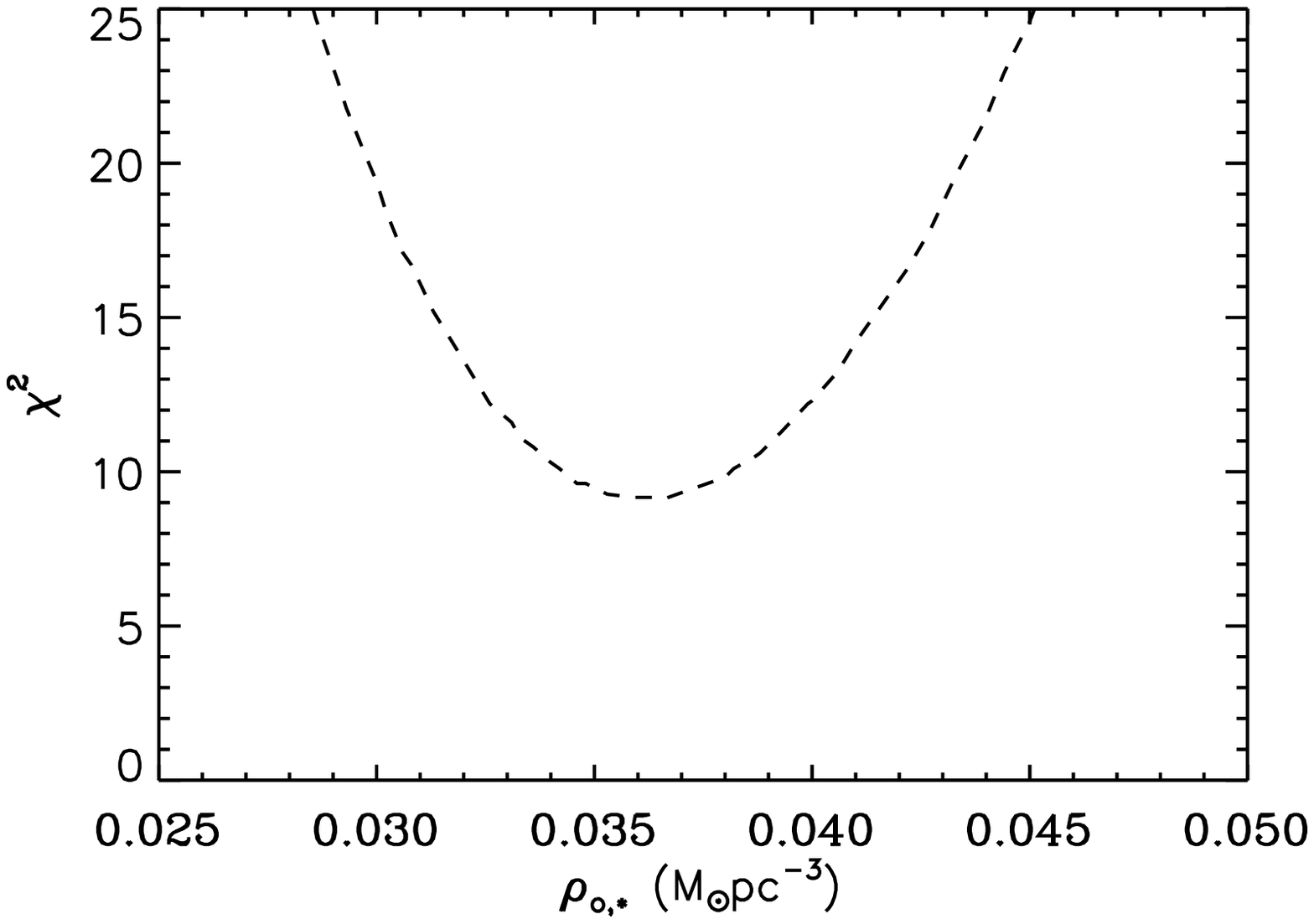}
\caption{\small{The lowest possible $\chi^2$ for Fornax in the case of MOND as a function of velocity anisotropy (left panel) and central density(right panel).}}
\label{fig:forr}
\end{figure*}

Although the range in  $\rho_{DM,0}$ and $r_{DM}$ are quite large, the defined range in enclosed mass is not nearly as vast. In Fig~\ref{fig:ML} we plot the enclosed mass within 300~pc for each of our 5 dSphs and
compare with the analysis of \cite{strigari08}. One can see that
 our sample of five appears consistent with a constant value of $10^7\msun$ within 300~pc although Sextans falls a factor of two short.  We
also overplot the enclosed mass from fits made by Angus (2010; where the original losVDs of \citealt{walker07} were fitted with the same DM density used here) and the trend for the enclosed mass to decrease after interloper removal is obvious, except for Leo I and Sextans. The lower $M_{300}$ from \citealt{angusdm} for Leo I is a subtle effect of fitting the outer 4 data points (from \citealt{walker07}), for which a fairly radial $\beta$ along with a large core, but relatively low central density allow a superior fit. In comparison, our fit to the cleaned sample is more consistent with a smaller core and larger central density. This makes our $M_{300}$ larger, but at 400~pc our enclosed mass drops below his model as expected.

We emphasise that although the dSph may have different velocity anisotropies than presented here, these are the simplest models that can achieve good fits to the data. Interestingly, \cite{wolf09} used a technique which enables the measurement of the enclosed mass at a single radius (coincidentally the half-light radius) which is independent of velocity anisotropy. The half-light radius of Carina is close to 300~pc and their value is similar to ours. Finally, \cite{aden09} have shown the mass within 300~pc of the ultra low luminosity dSph Hercules is reduced from $0.72^{+0.51}_{-0.21}\times10^{7}\msun$ to $0.19^{+0.11}_{-0.08}\times10^{7}\msun$ using another interloper removal technique, making it incompatible with the common mass scale. For our sample, the $M_{300}$ of Carina is actually reduced to be consistent with $10^7\msun$, but the low mass of Sextans is exacerbated. Excluding Sextans, the other 4 dSphs are consistent at 1-$\sigma$ with $M_{300}=1.2\times10^7\msun$

\subsection{MOND}
\protect\label{sec:mond}

In \citet{angus08} - see also \citet{kos09} - the losVDs as measured by \citet{walker07} for all 8 classical dSphs were taken without further investigation of interlopers. A Jeans analysis was then performed for each dSph, in the context of Modified Newtonian Dynamics or MOND \citep{milgrom83a,milgrom95}.  The only difference in the Jeans equation for MOND is that the gravitational acceleration must be modified, the stellar density and velocity anisotropy behave in the standard fashion. In the standard paradigm, the gravitational acceleration of a dSph is the combination of the weak stellar component and the dominant dark matter component. On the other hand, for MOND we must account for both the internal dynamics of the stellar component and the external gravitational field of the Milky Way which determine the MONDian gravity as a function of radius in the dSph. 

The formulation we use to define $g(r)$ for equation~(\ref{eqn:jeans}) is
$\mu(x)g(r)=g_*(r)$ where $x={\sqrt{g(r)2+g_{MW}(R_{MW})2} \over a_o}$.
Here, $a_o=1.2\times 10^{-10} ms^{-2}$ is the acceleration constant of
MOND, $\mu={x \over 1+x}$ is the MOND interpolating function we choose and
$g_{MW}(R_{MW})=(170~\kms)^{2}/R_{MW}$ is the external gravity of the Milky
Way at the dSph's Galactocentric radius assuming the terminal value of the
Milky Way's rotation speed to be $170~\kms$.

Since MOND does not benefit from the additional freedom of a dark matter halo and has an extra constraint due to the Milky Way's potential, the velocity anisotropy becomes a fitting function that enables one to uniquely determine the MONDian dynamical mass. Therefore, by fitting velocity anisotropy profiles to the losVDs, as well as accounting for the so called external field effect, \citet{angus08} was able to deduce the necessary $M/L$ of each dSph by comparing the dynamical mass with the luminosity.

From here, the required $M/L$ was compared with the expectation from models of stellar evolution given the star formation history of the dSphs in question. It was found that while a sensible range would be a V-band $M/L$ of 1-3, that only half (four out of eight) of the dSphs resided in this range (interestingly they are the most distant and most luminous dSphs), whereas a further two (Carina and Ursa Minor) had $M/L$s that were high, but consistent at 1-$\sigma$ and Draco and Sextans had exorbitantly high $M/L$s ($>$10). It was suggested that these high $M/L$s could be the result of tidal interactions with the Milky Way producing interloper stars in the foreground and background of the problematic dSphs.

\begin{figure*}
\centering
\includegraphics[angle=0,scale=0.45,bb=20 14 651 439]{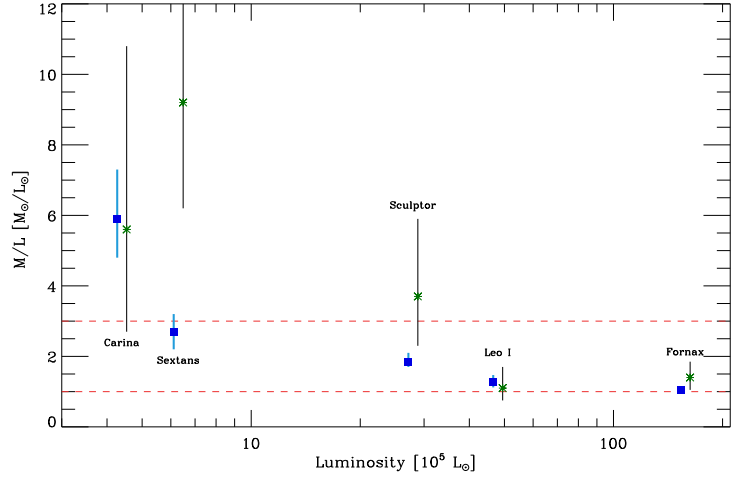}
\vspace{-0.2cm}
\caption{\small{Inferred $M/L$s in MOND using the old data (green asterisks, \citealt{angus08}) and our cleaned samples (blue squares). $M/L$s consistent with population synthesis models exist between the red dashed lines. The points referring to the same dSph are slightly shifted along the L axis for clarity.}}

\label{fig:MLmond}
\end{figure*}

Following exactly the same method, we have re-analysed the dSphs after
removing the interloper stars and have found that it leaves all three more luminous dSphs (Fornax, Leo I and Sculptor) with
appropriate $M/L$s (Figure \ref{fig:MLmond}), Sextans has a borderline $M/L$ (2.7$\pm$0.5), but the least luminous dSph Carina still has a suspiciously high $M/L$ (5.9$^{+1.4}_{-1.1}$).

The removal of interlopers by the caustic technique does not fully relieve MOND of its problem with the high $M/L$ of the low luminosity dSphs, regardless of the velocity anisotropy used (best fits given in Table 2).  Even more severe caustics cannot prune enough stars to give Carina a significantly lower $M/L$. As can be seen in Table 2, the MOND $M/L$ for Carina from the 3 other caustics we tried give 4.7 and 5.9. For the low luminosity dSphs to be consistent with MOND, they must either form from unusual initial mass functions or be tidally heated by their proximity to the Milky Way.
A crucial point to grasp here is that the idea of dSph tidal heating is
unusual in terms of the DM paradigm because tidal forces are too feeble at their
current locations. Although there are numerical studies of tidal harassment of dSphs by e.g. \cite{penarrubia0x}, \cite{klimentowski08} 
and \cite{lokas09}, they all rely on the dSphs plunging to pericentres that are a factor of three or four nearer the Milky Way 
($\approx 20$ kpc) than their current positions and observed proper motions allow 
(\citealt{piatek06,piatek07,piatek08}). Given the observed disk of satellites (\citealt{kroupa05}) and that the Magellanic Clouds are not bound to the Milky Way (\citealt{kallivayalil06a,kallivayalil06b}) these orbits are improbable, so other mechanisms should be investigated.

Furthermore, in the DM paradigm, it is not obvious how stars can be easily stripped
out of the dSph (although detailed studies are being made, e.g. \citealt{wetzel09}), because the DM halo protects the dSph from
tidal disruption, especially at the current distances of the dSphs. On the
other hand, in MOND the dSphs are more loosely bound - than their Newtonian counterparts with dark matter - and the internal dynamics vary with
distance from the Milky Way (due to the so-called external field effect, 
e.g. \citealt{milgrom83a,milgrom95,brada00,angus08}). When this external gravity (due to the Milky Way) reaches a
similar magnitude to the internal gravity of the dSph, the internal dSph gravity
decreases towards the simple Newtonian gravity of the stars and the dwarf and its velocities contract. For this reason, dSphs are more susceptible to tidal disruption in
MOND. This idea is currently being
studied with high-resolution N-body simulations in MOND (Angus et al. in preparation). Another idea to bear in mind is whether bottom heavy initial mass functions \citep[see][]{mie08} can give rise to the high $M/L$ ratios. One would have to explain why only the currently nearby, low luminosity dSphs are affected by this.

\begin{figure*}
\centering
\makebox[10cm][c]{\hspace{1.5cm}\textbf{Carina}\hspace{7.5cm}\textbf{Sextans}}
\includegraphics[angle=0,scale=.5,bb=20 20 506 277]{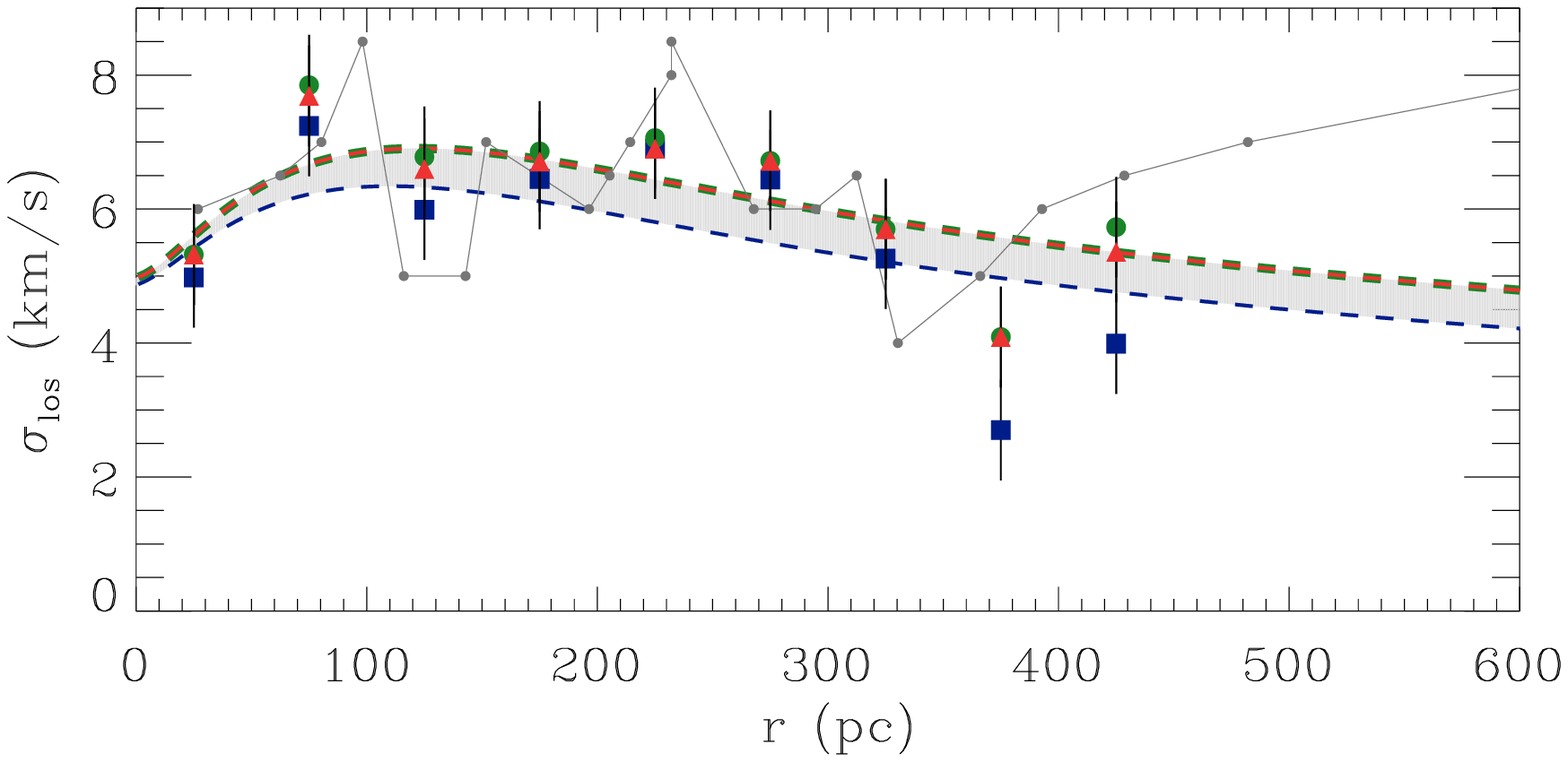}
\includegraphics[angle=0,scale=.5,bb=20 20 506 277]{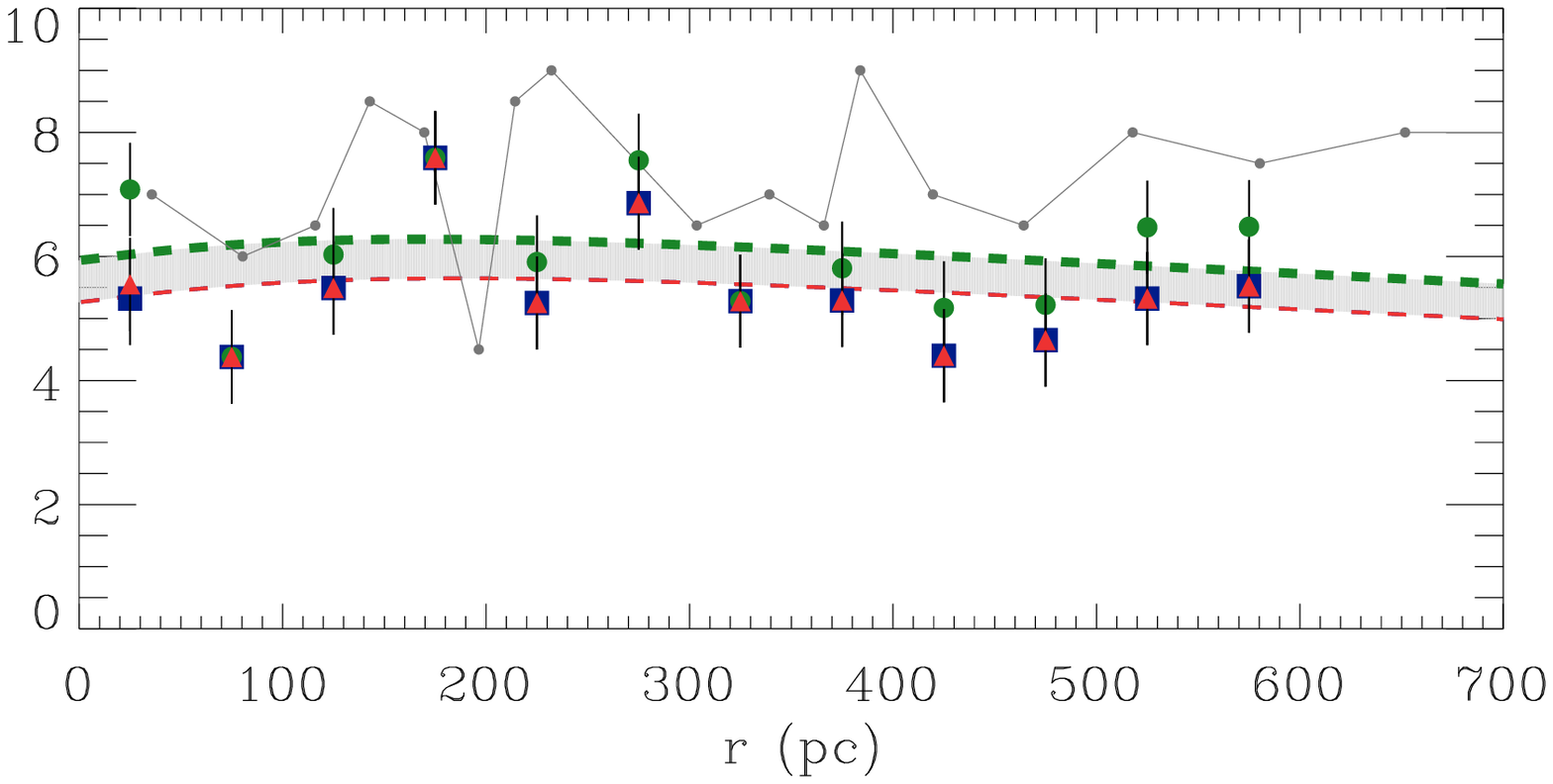}
\makebox[10cm][c]{\hspace{1.5cm}\textbf{Sculptor}\hspace{7.5cm}\textbf{Leo I}}
\includegraphics[angle=0,scale=.5,bb=20 20 506 277]{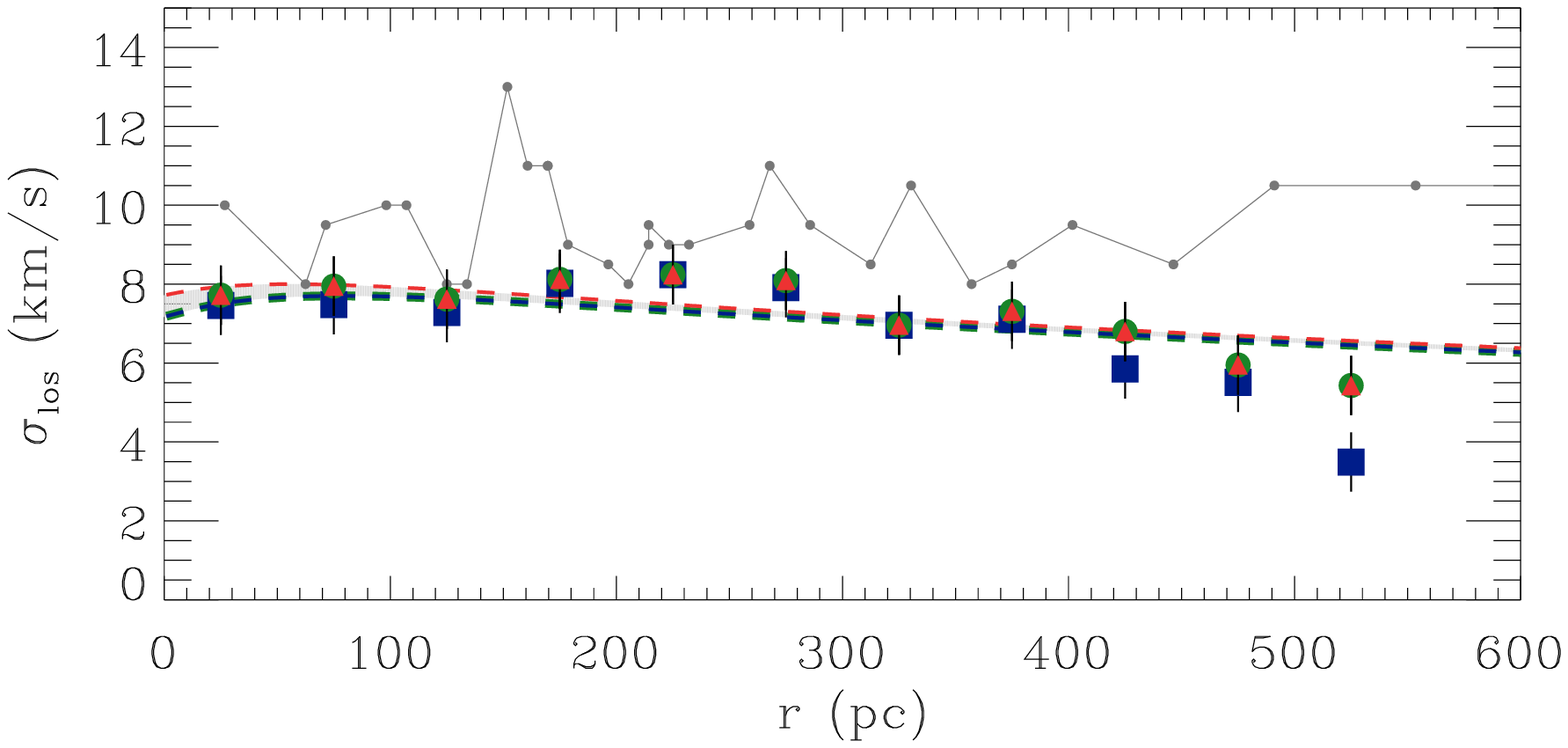}
\includegraphics[angle=0,scale=.5,bb=20 20 506 277]{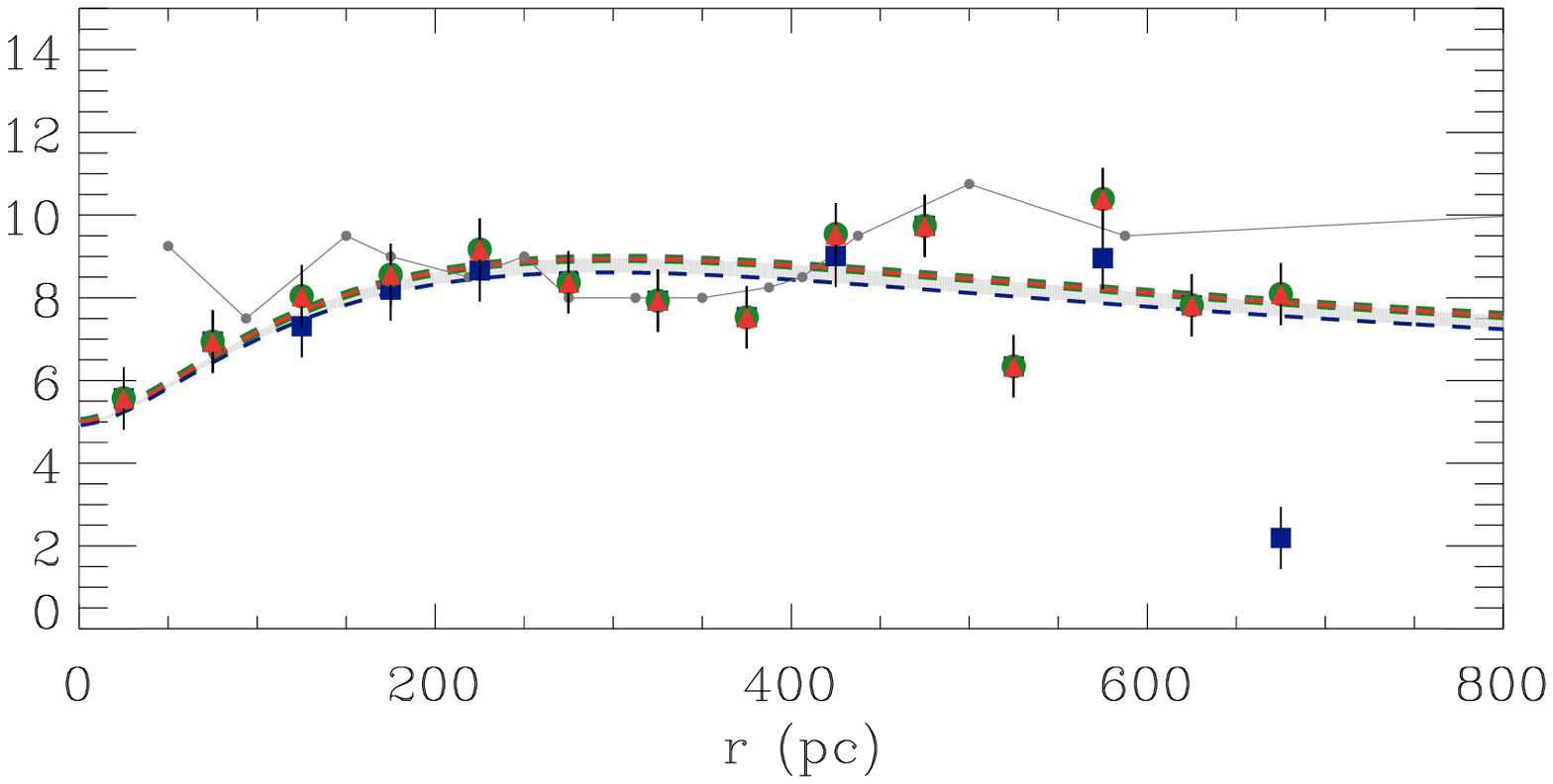}
\makebox[10cm][c]{\hspace{1.5cm}\textbf{Fornax}}
\includegraphics[angle=0,scale=.5,bb=20 20 506 277]{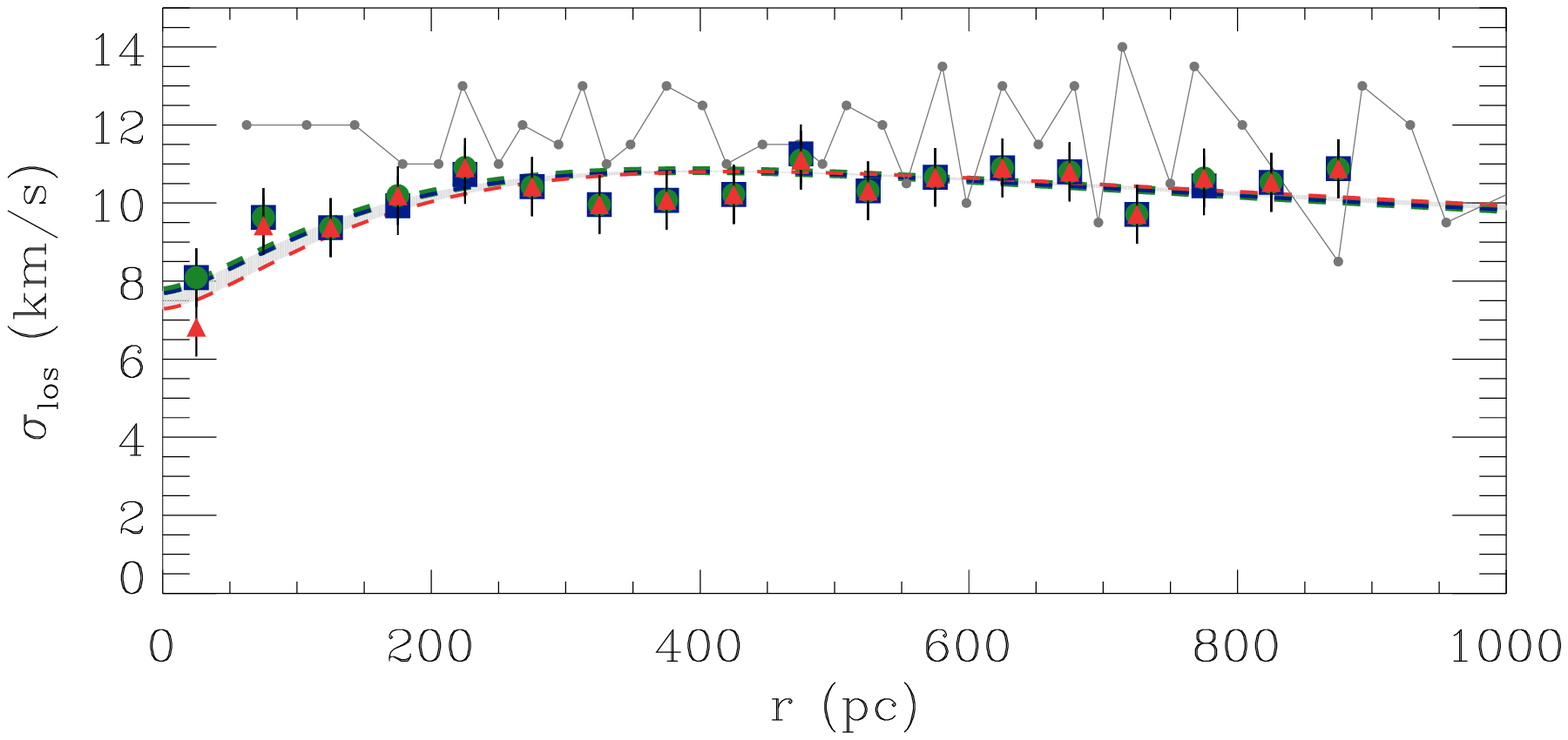}

\caption{\small{A comparison between the los velocity dispersions found using different star masses in the caustic technique as discussed in \S\ref{sec:starmass}. The red triangles corresponding to the intermediate caustics are supplemented by blue squares and green circles in each radius bin that signify the two other, extremal sets of caustics. The best MOND fit is given to each series of data points.}} 
\label{fig:veldisp_caus}
\end{figure*}

\begin{figure}
\includegraphics[angle=0,scale=.5,bb=60 40 506 277]{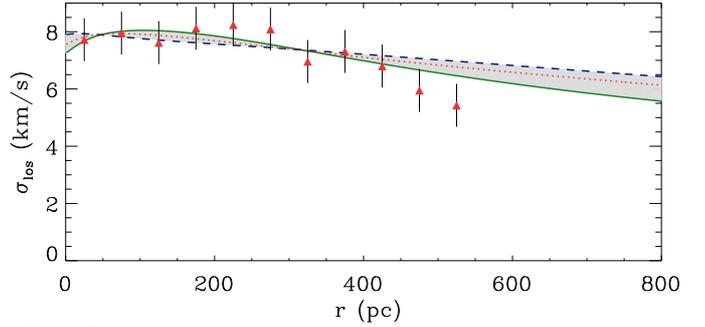}
\caption{\small{Fit to the line of sight velocity dispersion of Sculptor in MOND with different velocity anisotropy models, as discussed in \S\ref{sec:va}. The dotted blue line is for the best fit constant velocity anisotropy, $\beta=-0.03$ ($\chi^2$=5), the dashed red line is for a conservative (because an upper limit is imposed upon $\beta$) radially varying -0.13+${0.4 r^2\over {r^2+(300pc)^2}}$ ($\chi^2=3.75$), and the solid green line is for an unconstrained $\beta=-0.21+{0.9 r^2\over {r^2+(350pc)^2}}$ ($\chi^2=2.6)$.}} 
\label{sclbetinf}
\end{figure}

\begin{table*}
\centering
\begin{tabular}{ccccccc}
dSph & $M_{300}$ [$\times10^{6}\,\msun$] & $\rho_{DM,o}$ [$\msun pc^{-3}$] & $r_{DM}$ [pc]&$\beta_{o}$ & $\chi^2$ & n\\
Carina         & $8.8^{+3.5}$   &$3.72_{-2.27}$   &$50^{+115}$       &  $-3^{+2.57} $ & 5.8 & 9 \\
Fornax         & $11.7^{+6.2}_{-3.4}$   &$0.144_{-0.058}^{+0.206}$   &$450^{+220}_{-190} $    &  $-0.6^{+0.39}_{-0.8} $ & 8.8 &18\\
Sculptor         & $13.4_{-1.3}$   &$1.8_{-1.33}^{+6.05}$   &$83^{+84}_{-41} $      &  $-0.6^{+0.839}_{-2.4}$  & 5.1 (5.9)  &11\\
Sextans     & $5.5^{+1.3}_{-3.2}$   &$0.176_{-0.151}^{+21.8}$   &$179^{+402}_{-157} $      &  $-0.6^{+0.7}_{-2.4}$ & 14.1 (16.0)  &12\\
Leo I         & $12.9^{+2.7}_{-3.0}$   &$0.191_{-0.084}^{+0.43}$   &$347^{+253}_{-190} $      &  $-0.8^{+0.45}_{-2.2}$  & 24.8 (27.8) & 14  \\

\end{tabular}
\caption{List of the parameters used in fitting the dark halos to the cleaned los velocity dispersions. We give the enclosed masses at 300~pc and unreduced $\chi^2$ statistics for our fits. As discussed in \S 3.1 the minimum $\beta$ allowed is -3, and therefore all errors on one side are bound by this constraint (except in the case of Fornax). For instance, the lower error bar on $\beta$ reaches in most cases -3, which is not the 1-$\sigma$ error bar but the error bar at the $\chi^2$ in the brackets. The same applies to $\rho_{DM,o}$ and  $r_{DM}$. Related to this, the one sided error bars for Carina arise because the best fit is found at $\beta=-3$.}
\label{table1}
\end{table*} 

\begin{table*}
\centering
\begin{tabular}{ccccccccc}
dSph    & $R_{\rm{MW}}$ [kpc]  & $L_v$ [$\times10^5\lsun$] & $M/L$      & $\rho_{*,o}$ [$\times10^{-2}\msun pc^{-3}$] &$\beta(r)$ [pc]& $\chi^2$& n\\
Carina     & 101$\pm$5   & 4.4   & 5.9$^{+1.4}_{-1.1}$ (4.7,5.9) 	&13.8$^{+3.0}_{-2.7}$ 	& -0.75$^{+0.38}_{-0.5}$  & 7.2 &9 \\
Fornax     & 138$\pm$8   & 158   & 1.05$^{+0.09}_{-0.08}$ (1.05,1.07)  &3.6$\pm$0.3     	& -0.86$^{+0.2}_{-0.24}$   & 9.1   & 18  \\
Leo I      & 257$\pm$8   & 48    & 1.28$^{+0.19}_{-0.16}$ (1.12,1.13)  &5.14$^{+0.76}_{-0.64}$ & -1.5$^{+0.4}_{-0.5}$   & 25.7       & 14  \\
Sculptor   & 87$\pm$4    & 28    & 1.85$^{+0.25}_{-0.15}$ (1.71,1.85)  &22.0$\pm$ 3.0		& -0.13$^{+0.2}_{-0.27}$+${0.4 r^2\over {r^2+(300pc)^2}}$   &6.3 & 11 \\
Sextans    & 95$\pm$4    & 6.3   & 2.7$\pm$0.5 (2.7,3.3)  		&0.756$^{+0.1}_{-0.09}$ & -0.16$^{+0.2}_{-0.28}$   & 15.3 & 12  \\

\end{tabular}

\caption{Distance from the Milky Way to each dwarf, luminosity and list of the parameters used in fitting the MOND models to the cleaned los velocity dispersions. 
All velocity anisotropies are constant except Sculptor which is parametrised by $\beta(r)=\beta_o+{\beta_{\infty}r^2 \over r^2+r_{\beta}^2}$. We found no significant decrease of $\chi^2$ for the other 4 dSphs by giving extra freedom to the velocity anisotropy. The values between parentheses in the M/L column are the M/Ls obtained from the widest and narrowest caustics.}
\label{table2}
\end{table*}

\subsection{Star masses}
\protect\label{sec:starmass}
The star mass mentioned in \S\ref{sec:caus} is the only free parameter included in the caustic technique. To estimate the response of the caustics to changes in the star mass, we produced three sets of caustics for each dSph by setting three different values of the star mass (as shown in Figure \ref{fig:veldisp_caus}). The highest and lowest values contain the range in which we find a recognisable $\sigma$ plateau. The explicit values of the stellar mass used for each dwarf are given in Table~\ref{table3}.

The differences in the three sets of caustics for each dSph are not significant (Figure \ref{fig:caustics}, Table \ref{table4}), and the number 
of stars lying within the widest caustics and outside the narrowest one do not cause a 
large variation in the losVD profiles. This emphasises that the method 
is robust. Nevertheless, to identify the spread in the context of the mass models, we fitted models to the losVD 
from each caustic set for all 5 dSphs specifically in the case of MOND because differences in the $M/L$ are more obvious since the range is relatively tight. The three models used for Sculptor, Fornax and Leo are barely distinguishable (see $M/L$ column in Table 2) and yield MOND $M/L$s that differ from the main caustics at a much lower level than the errors we find at 1-$\sigma$. For Carina and Sextans the differences are moderate, but still lower than the 1-$\sigma$ errors.

\begin{table}
\centering
\begin{tabular}{c|ccccc}
dSph                &\multicolumn{3}{c}{star mass [$\times10^4\msun$]}  \\
                \cline{2-4}
               & $m_1$ & $m_{1.5}$ & $m_2$        \\
\hline
Carina         & 10    & 110      & 210  \\
Fornax         & 35    & 55       & 75   \\
Sculptor       & 15    & 55       & 95   \\
Sextans        & 10    & 30       & 50   \\
Leo I          & 30    & 665      & 1300 \\
\end{tabular}
\caption{Values of the stellar mass adopted for each dwarf that define the three sets of caustics for each dSph. The intermediate stellar mass $m_{1.5}$ is the one used for the general calculations.}
\label{table3}
\end{table} 

\begin{table}
\centering
\begin{tabular}{c|c|cccc}
dSph            & \multicolumn{4}{c}{number of galaxies} \\
\hline \hline            
                & catalogue &\multicolumn{3}{c}{within the caustics} \\
                             \cline{3-5}
                &           & $m_1$ & $m_{1.5}$ & $m_2$        \\
\hline
Carina          & 1981      & 638   & 665       & 674  \\
Fornax          & 2633      & 2274  & 2303      & 2308 \\
Sculptor        & 1541      & 1140  & 1176      & 1173 \\
Sextans         & 947       & 330   & 332       & 358  \\
Leo I           & 387       & 284   & 294       & 294  \\
\end{tabular}
\caption{Number of galaxies in the catalogue (from \citet{walker09} and within the three sets of caustics for each dwarf.}
\label{table4}
\end{table} 

\subsection{Velocity Anisotropy}
\protect\label{sec:va}
The velocity anisotropies we initially tested were generalised Osipkov-Merritt models that allow for non-isotropic orbits at small radii

\begin{equation}
\label{eqn:beta}
\beta(r)=\beta_0+\beta_{\infty}{1 \over 1 + \left({r_{\beta} \over r}\right)^2},
\end{equation}
where $\beta_{0}$, $\beta_{\infty}$ and $r_{\beta}$ are free parameters (although -$\infty<\beta<1$).

We found no preference for these models over constant velocity anisotropy, except in the case of Sculptor with MOND. In both cases (MOND and DM halos with our proposed density law), the tangential orbits (more negative $\beta$) at larger radii are necessary to fit the shape (in particular the flatness) of the losVD profile. The weightings of the tangential orbits are mild in the DM case (usually anywhere between $\beta=-3$ and $-0.5$) because of the freedom given by the DM halo. In MOND Fornax, Carina and Leo I prefer $\beta < -0.5$ (but only Leo I prefers $\beta < -1.0$). On the other hand, Sextans prefers almost isotropic orbits.

Sculptor definitely prefers a velocity anisotropy that not only varies with radius, but actually increases from very mildly tangential orbits at the centre ($\beta_o=-0.13$) to quite strongly radial orbits by the edge ($\beta_{\infty}=0.4$). In fact, the $\chi^2$ actually decreases slightly for even more radial orbits, but we imposed a maximum $\beta_{\infty}$ of 0.4. The reason for this (demonstrated in Fig~\ref{sclbetinf}) is to keep the velocity anisotropies from being abnormally radial, especially since the other dSphs prefer negative ones. In comparison, constant $\beta$ (with $\beta=-0.03$) has a $\chi^2$ of 5, $\beta=-0.13+{0.4 r^2\over {r^2+(220pc)^2}}$ has a $\chi^2$=3.75 and  $\beta=-0.21+{0.9 r^2\over {r^2+(350pc)^2}}$ has a $\chi^2$=2.6.

Perhaps the reason for Sculptor's odd losVD, and hence different velocity anisotropy, {\it may} be related to the reason the low luminosity dSphs have high $M/L$s. Although Sculptor is relatively luminous, it is also much nearer the Milky Way ($\sim$ 87~kpc) than Fornax and Leo I (138 and 257 kpc respectively) making it potentially more susceptible to tidal effects. However, this can only be tested with numerical simulations.

\cite{lokas09}, who used \cite{klimentowski08}'s technique to remove interloper stars from the catalogues of \cite{walker09} that we also use here, found that reasonable mass-follows-light models (no dark halo is assumed, simply the stars are given arbitrary $M/L$s) with constant velocity anisotropies could be fitted to the los VDs of the cleaned sample and their kurtoses. Interestingly, the losVDs emerging from the interloper removal technique used by \cite{lokas09} have a general trend to decline with radius, which allows for the good fits with constant velocity anisotropies and mass-follows-light models. On the other hand, four of our five (excluding Sculptor) stay very constant until the last measured point, meaning choice of interloper removal technique is key to constraining which models can be fitted to the data. Nevertheless, our ranges of $\beta$ are still very similar to \cite{lokas09}'s, with the exception of Carina. Our larger ranges in the allowed $\beta$s stem from the fact that we use dark halos instead of mass-follows-light models.
 
\section{Conclusion}

We have applied the caustic technique to remove star interlopers in the 5 classical dSphs 
observed by \cite{walker07,walker09} and \cite{mateo07}. We have applied the technique to each dSph, taking into account the systematics introduced by the free parameter in the binary tree construction, namely the star mass. We have found that there are no major discrepancies between the caustics calculated from each star mass (as shown by Figure \ref{fig:caustics} and the Tables \ref{table1} and \ref{table2}) in the mass range where a $\sigma$ plateau appears. Consequently, the velocity dispersion profiles are not strongly dependent on this
parameter of the caustic technique.

Using this technique we have shown that unbound/non-equilibrium stars are reasonably common in the low luminosity dSphs 
of Carina, Sextans and the relatively nearby (but more luminous) Sculptor. Removing these interlopers still leaves enclosed dark halo masses within 300~pc that appear to be constant and independent of the dSph luminosity \citep{strigari08,str09}, except for Sextans. 

We also investigated the implications for MOND, for which original analysis by \cite{angus08} showed the $M/L$s of the high luminosity dSphs were in perfectly reasonable agreement with
stellar population synthesis models, 
but the low luminosity dSphs had abnormally high $M/L$s. In particular, Sextans and Draco cast serious doubt on MOND's validity. After removing the interloper stars identified by the caustic technique, Sextans' $M/L$ has dropped to 2.7$\pm$0.5, but Carina still has a rather high $M/L$ of 5.9 $^{+1.4}_{-1.1}$ in
solar units. On the other hand, Fornax, Leo I and Sculptor were found to have very similar $M/L$s - 
1.05 $^{+0.09}_{-0.08}$, 1.28 $^{+0.19}_{-0.16}$, 1.85 $^{+0.25}_{-0.15}$, respectively. Further analysis of the final data release for Draco and Ursa Minor by \cite{walker07}, and MONDian N-body simulations will help to evaluate this incongruity.

The agreement between our results and previous identifications of the dSph interlopers based 
on kinematic and/or spectroscopic information show that the caustic technique can be applied 
with success to both galaxies and clusters of galaxies. The disadvantage of this technique
is the large number of stars or galaxies required to obtain robust estimates of the 
caustic location. However, the independence of the method on the dynamical state of
the system and the lack of any assumption on the underlying distribution of the 
member velocities provide substantial advantages over other widely used techniques
for removing interlopers.

\section*{ACKNOWLEDGMENTS}
GWA's research is supported by Universit\`a di Torino and Regione Piemonte. We 
also acknowledge partial support from the INFN grant PD51. 

\bibliographystyle{aa}

\end{document}